\documentclass[iop,apj]{emulateapj}

\usepackage{apjfonts}
\usepackage{psfig}
\usepackage{graphicx}

\usepackage[squaren,Gray]{SIunits}

\renewcommand{\tilde}{$\raise.17ex\hbox{$\scriptstyle\mathtt{\sim}$}$}

\def\aap{A\&A}

\def\apjl{ApJ}
\def\apjs{ApJS}

\def\arcsec{$^{\prime\prime}$}
\def\arcmin{$^\prime$}

\def\lir{$L_{\rm IR}$}

\def\l1.4{$L_{\rm 1.4GHz}$} \def\s1.4{$S_{\rm 1.4GHz}$}

\def\Md{$M_{\rm dust}$}
\def\Td{$T_{\rm dust}$}

\def\gs{\mathrel{\raise0.35ex\hbox{$\scriptstyle >$}\kern-0.6em
\lower0.40ex\hbox{{$\scriptstyle \sim$}}}}
\def\ls{\mathrel{\raise0.35ex\hbox{$\scriptstyle <$}\kern-0.6em
\lower0.40ex\hbox{{$\scriptstyle \sim$}}}}
\def\m@th{\mathsurround=0pt }
\def\eqalign#1{\null\,\vcenter{\openup1\jot \m@th
 \ialign{\strut\hfil$\displaystyle{##}$&$\displaystyle{{}##}$\hfil
 \crcr#1\crcr}}\,}

\begin{document}

\title{The space density of luminous dusty star-forming
  galaxies at $z>4$:\\
SCUBA-2 and LABOCA imaging of ultrared galaxies from {\it
    Herschel}-ATLAS}

\shorttitle{Submm imaging of ultrared galaxies}
\shortauthors{Ivison et al.}

\author{R.\,J.~Ivison\altaffilmark{1,2}}
\author{A.~\,J.~\,R.~Lewis\altaffilmark{2}}
\author{A.~Weiss\altaffilmark{3}}
\author{V.~Arumugam\altaffilmark{1,2}}
\author{J.\,M.~Simpson\altaffilmark{2}}
\author{W.\,S.~Holland\altaffilmark{4}}
\author{S.~Maddox\altaffilmark{2,5}}
\author{L.~Dunne\altaffilmark{2,5}}
\author{E.~Valiante\altaffilmark{5}}
\author{P.~van~der~Werf\altaffilmark{6}}
\author{A.~Omont\altaffilmark{7,8}}
\author{H.~Dannerbauer\altaffilmark{9,10,11}}
\author{Ian~Smail\altaffilmark{12}}
\author{F.~Bertoldi\altaffilmark{13}}
\author{M.~Bremer\altaffilmark{14}}
\author{R.\,S.~Bussmann\altaffilmark{15}}
\author{Z.-Y.~Cai\altaffilmark{16}}
\author{D.\,L.~Clements\altaffilmark{17}}
\author{A.~Cooray\altaffilmark{18}}
\author{G.~De~Zotti\altaffilmark{19,20}}
\author{S.\,A.~Eales\altaffilmark{5}}
\author{C.~Fuller\altaffilmark{5}}
\author{J.~Gonzalez-Nuevo\altaffilmark{19,21}}
\author{E.~Ibar\altaffilmark{22}}
\author{M.~Negrello\altaffilmark{5}}
\author{I.~Oteo\altaffilmark{2,1}}
\author{I.~P\'erez-Fournon\altaffilmark{9,10}}
\author{D.~Riechers\altaffilmark{15}}
\author{J.\,A.~Stevens\altaffilmark{23}}
\author{A.\,M.~Swinbank\altaffilmark{12}}
\author{J.~Wardlow\altaffilmark{12}}
\altaffiltext{1}{European Southern Observatory,
  Karl-Schwarzchild-Stra{\ss}e 2, D-85748 Garching, Germany}
\altaffiltext{2}{Institute for Astronomy, University of Edinburgh,
  Royal Observatory, Blackford Hill, Edinburgh EH9 3HJ, UK}
\altaffiltext{3}{Max-Planck-Institut f\"{u}r Radioastronomie,
  Auf dem H\"{u}gel 69, D-53121 Bonn, Germany}
\altaffiltext{4}{UK Astronomy Technology Centre, 
  Royal Observatory, Blackford Hill, Edinburgh EH9 3HJ, UK}
\altaffiltext{5}{School of Physics \& Astronomy, Cardiff University,
  Queen's Buildings, The Parade, Cardiff CF24 3AA, UK}
\altaffiltext{6}{Leiden Observatory, Leiden University, P.O.\ Box
  9513, NL-2300 RA Leiden, The Netherlands}
\altaffiltext{7}{UPMC Univ Paris 06, UMR 7095, IAP, 75014, Paris, France}
\altaffiltext{8}{CNRS, UMR7095, IAP, F-75014, Paris, France}
\altaffiltext{9}{IAC, E-38200 La  Laguna, Tenerife, Spain}
\altaffiltext{10}{Departamento de Astrofisica, Universidad de La
  Laguna, E-38205 La Laguna, Tenerife, Spain}
\altaffiltext{11}{Universit\"{a}t Wien, Institut f\"{u}r Astrophysik,
  T\"{u}rkenschanzstr.\ 18, 1180 Wien, Austria}
\altaffiltext{12}{Centre for Extragalactic Astronomy, Department of
  Physics, Durham University, South Road, Durham DH1 3LE, UK}
\altaffiltext{13}{Argelander-Institute for Astronomy, Bonn University,
  Auf dem Huegel 71, 53121 Bonn, Germany}
\altaffiltext{14}{H.\,H.\ Wills Physics Laboratory, University of
  Bristol, Tyndall Avenue, Bristol BS8 1TL, UK}
\altaffiltext{15}{Astronomy Department, Cornell University, Ithaca, NY 14853}
\altaffiltext{16}{CAS Key Laboratory for Research in Galaxies and
  Cosmology, Department of Astronomy, University of Science and
  Technology of China, Hefei, Anhui 230026, China}
\altaffiltext{17}{Astrophysics Group, Imperial College London,
  Blackett Laboratory, Prince Consort Road, London SW7 2AZ, UK}
\altaffiltext{18}{Department of Physics and Astronomy, University of
  California, Irvine, CA 92697, USA}
\altaffiltext{19}{SISSA, Via Bonomea 265, I-34136, Trieste, Italy}
\altaffiltext{20}{INAF-Osservatorio Astronomico di Padova, Vicolo
  dell’Osservatorio 5, I-35122 Padova, Italy}
\altaffiltext{21}{CSIC-UC, Avda los Castros s/n, 39005 Santander, Spain}
\altaffiltext{22}{Instituto de F\'{i}sica y Astronom\'{i}a,
  Universidad de Valpara\'{i}so, Avda.\ Gran Breta\~{n}a 1111,
  Valpara\'{i}so, Chile}
\altaffiltext{23}{Centre for Astrophysics, Science and Technology
  Research Institute, University of Hertfordshire, Hatfield AL10 9AB,
  UK}

\slugcomment{In press at The Astrophysical Journal}

\begin{abstract}
  Until recently, only a handful of dusty, star-forming galaxies
  (DSFGs) were known at $z>4$, most of them significantly amplified by
  gravitational lensing.  Here, we have increased the number of such
  DSFGs substantially, selecting galaxies from the uniquely wide 250-,
  350- and 500-$\mu$m {\it Herschel}-ATLAS imaging survey on the basis
  of their extremely red far-infrared colors and faint 350- and
  500-$\mu$m flux densities -- {\it ergo} they are expected to be
  largely unlensed, luminous, rare and very distant.  The addition of
  ground-based continuum photometry at longer wavelengths from the
  James Clerk Maxwell Telescope (JCMT) and the Atacama Pathfinder
  Experiment (APEX) allows us to identify the dust peak in their
  spectral energy distributions (SEDs), better constraining their
  redshifts.  We select the SED templates best able to determine
  photometric redshifts using a sample of 69 high-redshift, lensed
  DSFGs, then perform checks to assess the impact of the CMB on our
  technique, and to quantify the systematic uncertainty associated
  with our photometric redshifts, $\sigma=0.14\,(1+z)$, using a sample
  of 25 galaxies with spectroscopic redshifts, each consistent with
  our color selection.  For {\it Herschel}-selected ultrared galaxies
  with typical colors of $S_{500}/S_{250}\sim 2.2$ and
  $S_{500}/S_{350}\sim 1.3$ and flux densities,
  $S_{500}\sim 50$\,mJy, we determine a median redshift, $\hat{z}_{\rm
    phot}=3.66$, an interquartile redshift range, 3.30--4.27, with a
  median rest-frame 8--1000-$\mu$m luminosity, $\hat{L}_{\rm IR}$, of
  $1.3\times 10^{13}$\,L$_\odot$.  A third lie at $z>4$, suggesting a
  space density, $\rho_{z>4}$, of $\approx 6 \times
  10^{-7}$\,Mpc$^{-3}$.  Our sample contains the most luminous known
  star-forming galaxies, and the most over-dense cluster of
  starbursting proto-ellipticals yet found.
\end{abstract}


\section{Introduction}
\label{intro}

The first deep submillimeter (submm) imaging surveys -- made possible
by large, ground-based telescopes equipped with highly multiplexed
bolometer arrays \citep[e.g.][]{kreysa98,holland99} -- resolved a
previously unknown population of submm-bright galaxies, or dusty
star-forming galaxies \citep[hereafter DSFGs ---][]{smail97, barger98,
  hughes98}.  Interferometric imaging refined the positions of these
DSFGs sufficiently to allow conventional optical spectroscopic
observations, and they were then shown to lie at $z>1$
\citep[e.g.][]{chapman03}, and to be a thousand times more numerous
than their supposed local analogs, ultraluminous infrared (IR)
galaxies \citep[ULIRGs --- e.g.][]{sanders96}.

The Spectral and Photometric Imaging Receiver \citep[SPIRE
---][]{griffin10} on board {\it Herschel} \citep{pilbratt10} gave
astronomers a new tool to select dusty galaxies.  Moreover,
simultaneous imaging through three far-infrared filters at 250, 350
and 500\,$\mu$m enables the selection of `ultrared DSFGs' in the early
Universe, $z>4$. The space density and physical properties of the
highest-redshift starbursts provide some of the most stringent
constraints on galaxy-formation models, since these galaxies lie on
the most extreme tail of the galaxy stellar mass function
\citep[e.g.][]{hainline11}.

\citet{cox11} were the first to search amongst the so-called
`500-$\mu$m risers' ($S_{250} < S_{350} < S_{500}$, where
$S_{\lambda}$ is the flux density at $\lambda$\,$\mu$m), reporting
extensive follow-up observations of one of the brightest, reddest
DSFGs in the first few 16-deg$^2$ tiles of the $\approx 600$-deg$^2$
imaging survey, {\it H}-ATLAS \citep[{\it Herschel} Astrophysical
Terahertz Large Area Survey ---][]{eales10}, a lensed starburst at
$z=4.2$, G15.141 or HATLAS\,J142413.9$+$022304, whose clear,
asymmetric double-peaked CO lines betray an asymmetric disk or ring,
and/or the near-ubiquitous merger found in such systems
\citep[][]{engel10}.  \citet{dowell14} demonstrated the effectiveness
of a similar SPIRE color-selection technique, finding
1HERMES\,S350\,J170647.8$+$584623 at $z=6.3$ \citep{riechers13} in the
northern 7-deg$^2$ First Look Survey field \citep[see
also][]{asboth16}.  Meanwhile, relatively wide and shallow surveys
with the South Pole Telescope (SPT) have allowed the selection of
large numbers of gravitationally lensed DSFGs \citep{vieira10}. These
tend to contain cold dust and/or to lie at high redshifts
\citep{vieira13,weiss13,strandet16}, due in part to their selection at
wavelengths beyond 1\,mm, which makes the survey less sensitive to
warmer sources at $z\approx 1$--3.

In this paper, we report efforts to substantially increase the number
of ultrared DSFGs, using a similar color-selection method to isolate
colder and/or most distant galaxies, at $z>4$, a redshift regime where
samples are currently dominated by galaxies selected in the rest-frame
ultraviolet \citep[e.g.][]{ellis13}.  Our goal here is to select
galaxies that are largely unlensed, rare and very distant, modulo the
growing optical depth to lensing at increasing redshift.  We hope to
find the progenitors of the most distant quasars, of which more than a
dozen are known to host massive ($>10^8$\,M$_\odot$) black holes at
$z>6$ \citep[e.g.][]{fan01, mortlock11}.  We would expect to find
several in an area the size of {\it H}-ATLAS, $\approx 600$\,deg$^2$,
if the duration of their starburst phase is commensurate with their
time spent as `naked' quasars.  We accomplish this by searching over
the whole {\it H}-ATLAS survey area -- an order of magnitude more area
than the earlier work in {\it H}-ATLAS.

We exploit both ground- and space-based observations, concentrating
our efforts in a flux-density regime, $S_{\rm 500}<100$\,mJy where
most DSFGs are not expected to be boosted significantly by
gravitational lensing \citep{negrello10,conley11}.  We do this
partly to avoid the uncertainties associated with lensing
magnification corrections and differential magnification
\citep[e.g.][]{serjeant12}, partly because the areal coverage of our
{\it Herschel} survey would otherwise yield only a handful of targets,
and partly because wider surveys with the SPT are better suited to
finding the brighter, distant, lensed population.

In the next section we describe our data acquisition and our methods
of data reduction.  We subsequently outline our sample selection
criteria before presenting, analyzing, interpreting and discussing our
findings in \S4.  Our conclusions are outlined in \S5.  Follow-up
spectral scans of a subset of these galaxies with the Atacama Large
Millimeter Array (ALMA) and with Institute Radioastronomie
Millimetrique's (IRAM's) Northern Extended Millimeter Array (NOEMA)
are presented by \citet{fudamoto16}.  Following the detailed ALMA
study by \citet{oteo16cplus} of one extraordinarily luminous DSFG from
this sample, \citet{oteo16hires} present high-resolution continuum
imaging of a substantial subset of our galaxies, determining the size
of their star-forming regions and assessing the fraction affected by
gravitational lensing.  Submillimeter imaging of the environments of
the reddest galaxies using the 12-m Atacama Pathfinder Telescope
(APEX) are presented by \citet{lewis16}.  A detailed study of a
cluster of starbursting proto-ellipticals centered on one of our
reddest DSFGs is presented by \citet{oteo16grh}.

We adopt a cosmology with $H_0 = 71$\,km\,s$^{-1}$\,Mpc$^{-1}$,
$\Omega_{\rm m}=0.27$ and $\Omega_\Lambda = 0.73$.

\section{Sample selection}
\label{sample}

\subsection{Far-infrared imaging}

We utilize images created for the {\it H}-ATLAS Data Release 1
\citep{valiante16}, covering three equatorial fields with right
ascensions of 9, 12 and 15\,hr, the so-called GAMA09, GAMA12 and
GAMA15 fields, each covering $\approx 54$\,deg$^2$; in the north, we
also have $\approx 170$\,deg$^2$ of areal coverage in the North
Galactic Pole (NGP) field; finally, in the south, we have $\approx
285$\,deg$^2$ in the South Galactic Pole (SGP) field, making a total
of $\approx 600$\,deg$^2$.  The acquisition and reduction of these
{\it Herschel} parallel-mode data from SPIRE and PACS
\citep[Photoconductor Array Camera and Spectrometer
---][]{poglitsch10} for {\it H}-ATLAS are described in detail by
\citet{valiante16}.  Summarising quickly: before the subtraction of a
smooth background or the application of a matched filter, as described
next in \S\ref{madx}, the 250-, 350- and 500-$\mu$m SPIRE maps
exploited here have 6, 8 and 12\arcsec\ pixels, point spread functions (PSFs)
with azimuthally-averaged {\sc fwhm} of 17.8, 24.0 and 35.2\arcsec\
and mean instrumental [confusion] r.m.s.\ noise levels of 9.4 [7.0],
9.2 [7.5] and 10.6 [7.2]\,mJy, respectively, where $\sigma_{\rm
  total}=\sqrt{\sigma^2_{\rm conf}+\sigma^2_{\rm instr}}$.

\subsection{Source detection}
\label{madx}

Sources were identified and flux densities were measured using a
modified version of the Multi-band Algorithm for source eXtraction
({\sc madx}; Maddox et al., in prep).  {\sc madx} first subtracted a
smooth background from the SPIRE maps, and then filtered them with a
`matched filter' appropriate for each band, designed to mitigate the
effects of confusion \citep[e.g.][]{chapin11}.  At this stage, the map
pixel distributions in each band have a highly non-Gaussian positive
tail because of the sources in the maps, as discussed at length for
the unfiltered maps by \citet{valiante16}.

Next, 2.2-$\sigma$ peaks were identified in the 250-$\mu$m map, and
`first-pass' flux-density estimates were obtained from the pixel
values at these positions in each SPIRE band. Sub-pixel positions were
estimated by fitting to the 250-$\mu$m peaks, then more accurate flux
densities were estimated using bi-cubic interpolation to these
improved positions. In each band, the sources were sorted in order of
decreasing flux density using the first-pass pixel values, and a
scaled PSF was subtracted from the map, leaving a residual map used to
estimate fluxes for any fainter sources. This step prevents the flux
densities of faint sources being overestimated when they lie near
brighter sources. In the modified version of {\sc madx}, the PSF
subtraction was applied only for sources with 250-$\mu$m peaks greater
than 3.2$\sigma$.  The resulting 250-$\mu$m-selected sources were
labelled as {\sc bandflag}=1 and the pixel distribution in the
residual 250-$\mu$m map is now close to Gaussian, since all of the
bright 250-$\mu$m sources have been subtracted.

The residual 350-$\mu$m map, in which the pixel distribution retains a
significant non-Gaussian positive excess, was then searched for
sources, using the same algorithms as for the initial 250-$\mu$m
selection.  Sources with peak significance more than 2.4-$\sigma$ in
the 350-$\mu$m residual map are saved as {\sc bandflag}=2 sources.
Next, the residual 500-$\mu$m map was searched for sources, and
2.0-$\sigma$ peaks are saved as {\sc bandflag}=3 sources.

Although the pixel distributions in the final 350- and 500-$\mu$m
residual images are much closer to Gaussian than the originals, a
significant non-Gaussian positive tail remains, due to subtracting
PSFs from sources that are not well fit by the PSF. Some of these are
multiple sources detected as a single blend, while some are extended
sources.  Since even a single, bright, extended source can leave
hundreds of pixels with large residuals --- comparable to the
residuals from multiple faint red sources --- it is not currently
feasible to disentangle the two. 

For the final catalogue, we keep sources only if they are above
3.5$\sigma$ in any one of the three SPIRE bands.  For each source, the
astrometric position was determined by the data in the initial
detection band.  No correction for flux boosting has been
applied\footnote{For our selection process this correction depends
  sensitively on the flux density distribution of the sources as well
  as on their colour distribution, neither of which is known well,
  such that the uncertainty in the correction is then larger than the
  correction itself (see also \S\ref{nzsummary}).}. The catalogue thus
created contains $7\times 10^5$ sources across the five fields
observed as part of H-ATLAS.

\subsection{Parent sample of ultrared DSFG candidates}
\label{parentsample}

\begin{figure}
\centerline{\psfig{file=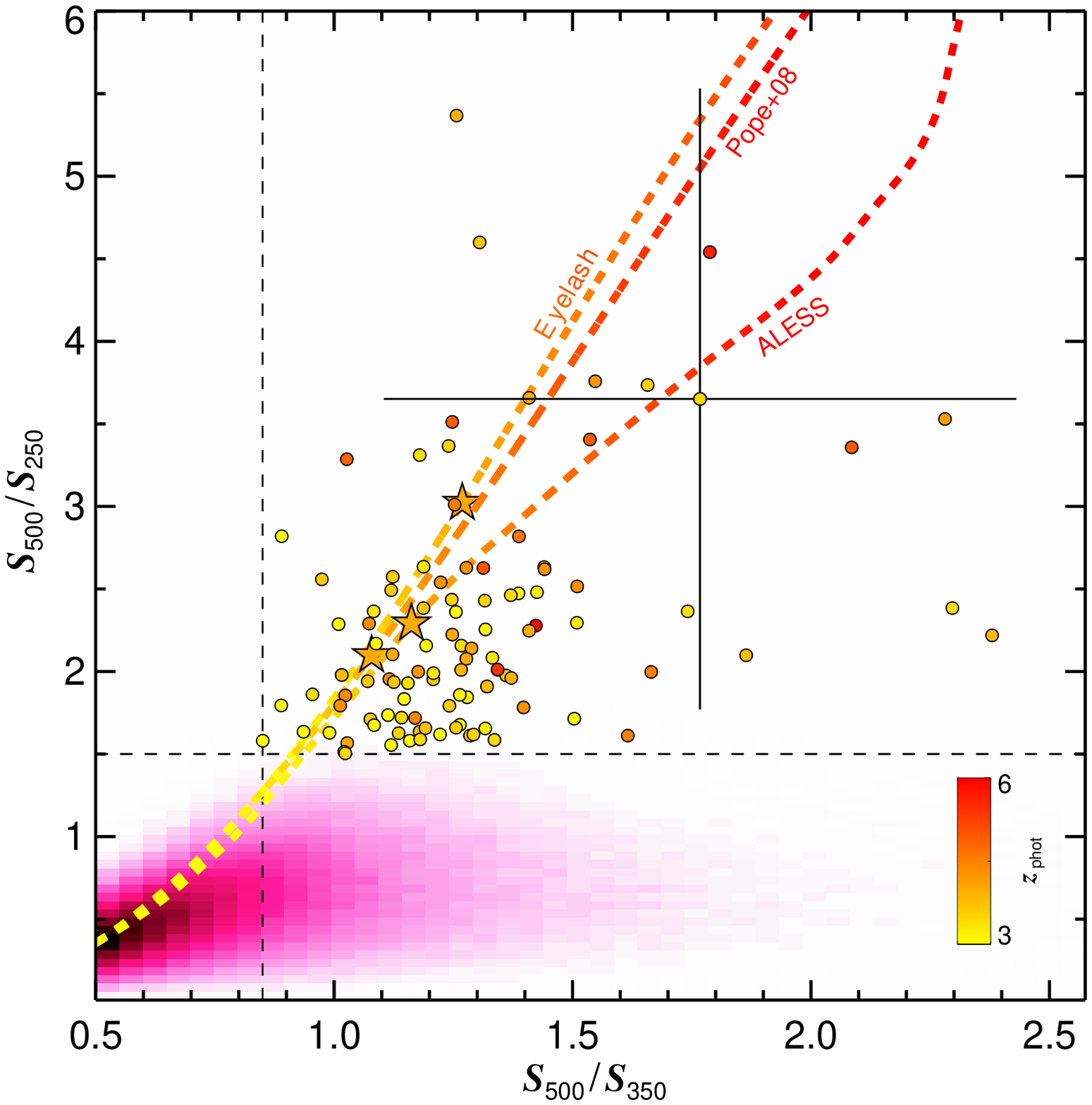,width=3.3in,angle=0}}
\caption{$S_{350}/S_{500}$ versus $S_{250}/S_{500}$ for our sample,
  overlaid with the redshift tracks expected for a galaxy with the SED
  of the Cosmic Eyelash \citep{swinbank10, ivison10fts} and for two
  SED templates that were synthesized for submm-selected DSFGs by
  \citet{pope08} and \citeauthor{swinbank14} (2014, ALESS). To match
  our color-selection criteria, galaxies must have $S_{500}/S_{250}\ge
  1.5$ and $S_{500}/S_{350}\ge 0.85$ and thus lie in the top-right
  region of the plot.  The points representing our sample (and the
  redshift track) are color-coded according to their photometric
  redshifts, as described in \S\ref{photoz}. The $z=4$ points on the
  redshift tracks are marked with orange stars.  A representative
  color uncertainty is shown.  Sources from the Phase 1 data release
  of {\it H}-ATLAS lie in the black-pink cloud \citep{valiante16}.}
\label{fig:colourcolour}
\end{figure}

\begin{figure}
\centerline{\psfig{file=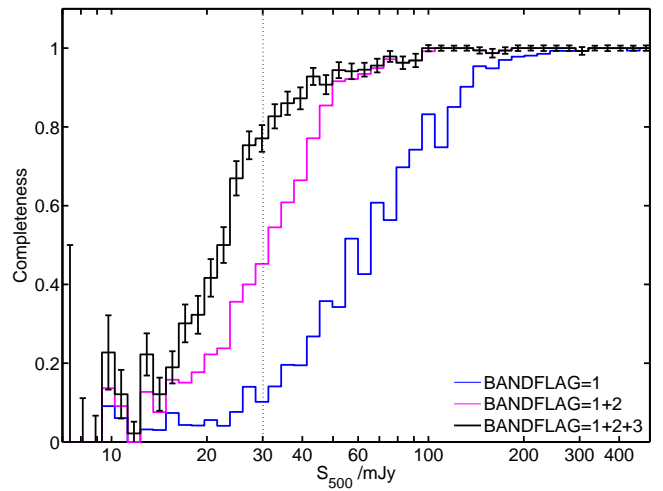,width=3.4in,angle=0}}
\caption{Completeness as a function of 500-$\mu$m flux density, as
  assessed by injecting fake sources with colors consistent with the
  SEDs of the ultrared DSFGs we expect to detect.  For the individual
  fake SPIRE images (see \S\ref{conventionalcompleteness}),
  completeness is consistent with expectations for sources at a given
  signal-to-noise ratio (SNR).  Using all three {\sc bandflag} values
  results in a relatively high level of completeness ($77\pm 3$\%) down
  to 30\,mJy, the flux-density level (marked with a dotted line) at
  which we have selected our sample.  Adding the {\sc bandflag}=2 and
  3 sources improves the completeness significantly.}
\label{fig:completeness}
\end{figure}
 
Definition of our target sample began with the 7,961 sources detected
at $\ge3.5$-$\sigma$ at 500\,$\mu$m, with $S_{500}/S_{250}\ge 1.5$ and
$S_{500}/S_{350}\ge 0.85$, as expected for DSFGs at $z\gs 4$
\citep[see the redshift tracks of typical DSFGs, e.g.\ the Cosmic
Eyelash, SMM\,J2135$-$0102 ---][in
Fig.~\ref{fig:colourcolour}]{swinbank10, ivison10fts} of which 29, 42
and 29\% are {\sc bandflag} = 1, 2 and 3, respectively.

\subsubsection{Conventional completeness}
\label{conventionalcompleteness}
 
To calculate the fraction of real, ultrared DSFGs excluded from the
parent sample because of our source detection procedures, we injected
15,000 fake, PSF-convolved point sources into our {\it H}-ATLAS images
\citep[following][]{valiante16} with colors corresponding to the
spectral energy distribution (SED) of a typical DSFG, the Cosmic
Eyelash, at redshifts between 0 and 10.  The mean colors of these fake
sources were $S_{500}/S_{250} = 2.25$ and $S_{500}/S_{350}= 1.16$,
cf.\ the median colors for the sample chosen for ground-based imaging
(\S\ref{eyeballing}), $S_{500}/S_{250} = 2.15$ and $S_{500}/S_{350}=
1.26$, so similar.  Values of $S_{250}$ were set to give a uniform
distribution in log$_{10}\,S_{250}$.  We then re-ran the same source
detection process described above (\S\ref{madx}), as for the real
data, matching the resulting catalog to the input fake catalog.

To determine the completeness for the ultrared sources, we have
examined how many of the recovered fake sources match our color
criteria, as a function of input $S_{500}$ and {\sc bandflag}.
Fig.~\ref{fig:completeness} shows how adding the {\sc bandflag}=2 and
3 sources improves the completeness: the blue line is for {\sc
  bandflag}=1 only; magenta is for {\sc bandflag}=1 and 2, and black
shows {\sc bandflag}=1, 2 and 3.  Selecting only at 250\,$\mu$m yields
a completeness of 80\% at 100\,mJy; including {\sc bandflag}=2 sources
pushes us down to 50\,mJy; using all three {\sc bandflag} values gets
us down to 30\,mJy.  We estimate a completeness at the flux-density
and color limits of the sample presented here of $77\pm 3$\%.

\begin{figure}
  \centerline{\psfig{file=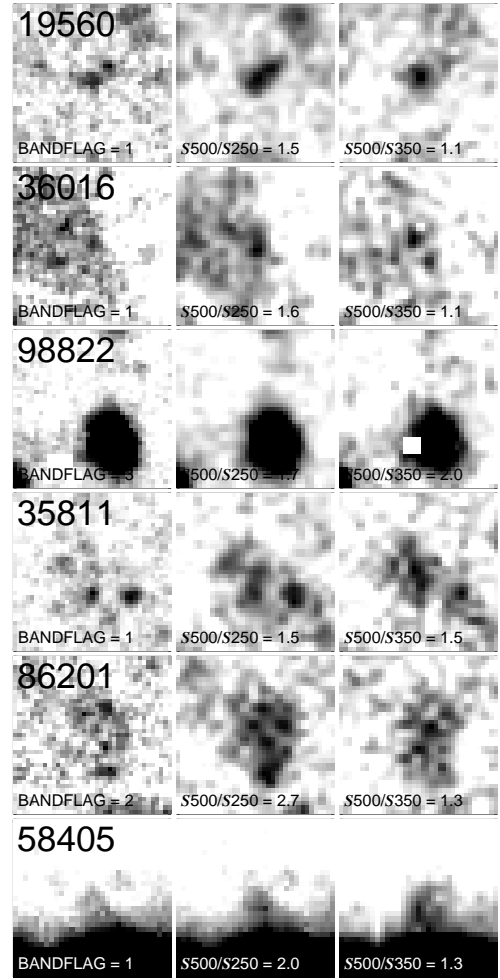,width=2.55in,angle=0}}
  \caption{{\it Herschel} SPIRE imaging of candidate ultrared DSFGs
    from our parent sample of 7,961 sources, each displayed from $-6$
    to +60\,mJy\,beam$^{-1}$, chosen to illustrate the different
    reasons that sources were excluded from the sample to be observed
    by SCUBA-2 and LABOCA by our eyeballing team.  In each column,
    from left to right, we show 250-, 350- and 500-$\mu$m cut-out
    images, each $3^\prime\times 3^\prime$ and centered on the
    (labelled) galaxy. The 250-$\mu$m cut-out images have been
    convolved with a 7$^{\prime\prime}$ Gaussian.  North is up and
    East is left. The field labelled 19560 is an example where
    emission from one or more 250-$\mu$m sources is missed or dealt
    with poorly by {\sc madx}, leading to misleading colors.  None of
    the candidates in this category are likely to be genuine, ultrared
    DSFGs.  The examples labelled 36016, 35811 and 86201 show confused
    regions in which the {\sc madx} flux densities and colors were
    judged unreliable.  We estimate that up to $\approx$55\% of these
    fields could contain genuine, ultrared DSFGs.  The bright galaxy
    in the field labelled 98822 has led to a spurious detection by
    {\sc madx} -- such examples are rare, fortunately, and {\sc madx}
    is in fact capable of identifying plausible ultrared DSFGs
    alongside very bright, local galaxies, as illustrated in the lower
    row for the field labelled 58405.}
\label{fig:examples}
\end{figure} 

\subsubsection{Eyeballing}
\label{eyeballing}

Of these sources, a subset of 2,725 were eyeballed by a team of five
(RJI, AJRL, VA, AO, HD) to find a reliable sub-sample for imaging with
SCUBA-2 and LABOCA.  As a result of this step, 708 ($26\pm 5$\%) of
the eyeballed sources were deemed suitable for ground-based follow-up
observations, where the uncertainty is taken to be the scatter amongst
the fractions determined by individual members of the eyeballing
team. Fig.~\ref{fig:examples} shows typical examples of the remainder
-- those not chosen\footnote{Fig.~\ref{fig:examples} also shows a case
  where {\sc madx} succeeds in cataloging an ultrared DSFG candidate
  that is nestled alongside a very bright, local galaxy.}  -- usually
because visual inspection revealed that blue (250-$\mu$m) emission had
been missed or underestimated by {\sc madx} (49\% of cases).  None of
these are likely to be genuine, ultrared DSFGs.  The next most common
reason for rejection (22\% of cases) was heavy confusion, such that
the assigned flux densities and colors were judged to be unreliable.
For the remaining 3\%, the 350- and/or 500-$\mu$m morphologies were
suggestive of Galactic cirrus or an imaging artifact.

\subsubsection{Completeness issues related to eyeballing}

Our team of eyeballers estimated that up to 14\% of the candidates
excluded by our eyeballing team -- i.e.\ up to 55\% of those in the
latter two categories discussed in \S\ref{eyeballing}, or plausibly
roughly half as many again as those deemed suitable for ground-based
follow-up observations -- could in fact be genuine, ultrared DSFGs.
Phrased another way, the procedure was judged to recover at least 64\%
of the genuine, ultrared DSFGs in the parent sample.

Without observing a significant subset of the parent sample with
SCUBA-2 or LABOCA, which would be prohibitively costly and
inefficient, it is not possible to know exactly what fraction of
genuine, ultrared DSFGs were missed because of our eyeballing
procedure.  However, it is possible to determine the fraction of
sources that were missed in a more quantitative manner than we have
accomplished thus far.  To do this, a sample of 500 fake, injected
ultrared sources -- with the same flux density and color distribution
as the initial sample -- were given to the same team of eyeballers for
classification, using the same criteria they had used previously,
along with the same number of real, ultrared DSFG candidates.  The
fraction of genuine, ultrared DSFGs accepted by the eyeballing team is
then taken to be the fraction of fake, injected sources assessed to be
worthy of follow-up observations during this eyeballing process:
$69\pm 8$\%, cf.\ at least 64\%, as estimated earlier by the
eyeballing team.

\subsection{Summary of issues affecting sample completeness}
\label{summarycompleteness}

Since we have faced a considerable number of completeness issues, it
is worth summarising their influence on our sample.

Based on robust simulations, we estimate that $\mathcal{C_{\rm MADX}}
=77\pm 3$\% of genuine, ultrared DSFGs made it through our {\sc madx}
cataloging procedures; of these, we eyeballed $\mathcal{C_{\rm
    eye}}=34$\%, of which $26\pm 5$\% were deemed suitable for
follow-up observations with SCUBA-2 and/or LABOCA by our eyeballing
team.  A final set of simulations suggest that the eyeballing process
was able to recover $\mathcal{C_{\rm check}} = 69\pm 8$\% of the
available ultrared DSFG population from the parent {\sc madx} catalog.

Of those selected for further study, a random subset of 109 were
observed with SCUBA-2 and/or LABOCA (\S\ref{observations}), just over
$\mathcal{C_{\rm obs}} =15$\% of the sample available from our
eyeballing team. Their SPIRE colors are shown in
Fig.~\ref{fig:colourcolour}. The {\sc bandflag} $=1$, 2 and 3 subsets
make up 48, 53 and 8 of this final sample, respectively.

To estimate the number of $z>4$ DSFGs across our survey fields,
detectable to $S_{500}>30$\,mJy with $S_{500}/S_{250}\ge 1.5$ and
$S_{500}/S_{350}\ge 0.85$, we must scale up the number of $z>4$ DSFGs
found amongst these 109 targets by $\mathcal{C_{\rm MADX}} \times
\mathcal{C_{\rm eye}} \times \mathcal{C_{\rm check}} \times
\mathcal{C_{\rm obs}})^{-1}=36.0\pm 8.2$, where we have included (in
quadrature) the uncertainty in the fraction deemed suitable for
follow-up observations with SCUBA-2 and/or LABOCA.  In a more
conventional sense, the completeness, $\mathcal{C} = 0.028\pm 0.006$.

We note that although we are unable to satisfactorily quantify the
number of DSFGs scattered by noise from the cloud shown in the
bottom-left corner of Fig.~\ref{fig:colourcolour} into our ultrared
DSFG color regime, these DSFGs will be amongst the fraction shown to
lie at $z_{\rm phot}<4$ (\S\ref{photoz}) and so a further correction
to the space density of $z>4$ DSFGs (\S\ref{spacedensity}) is not
required.

\section{Submm observations and data reduction}
\label{observations}

\subsection{850-$\mu$m continuum imaging with SCUBA-2}
\label{scuba2obs}

Observations of 109 ultrared DSFGs were obtained using SCUBA-2
\citep{holland13}, scheduled flexibly during the period 2012--13, in
good or excellent weather. The precipitable water vapor (PWV) was in
the range 0.6--2.0\,mm, corresponding to zenith atmospheric opacities
of $\approx 0.2$--0.4 in the SCUBA-2 filter centered at 850\,$\mu$m
with a passband width to half power of 85\,$\mu$m.  The {\sc fwhm} of
the main beam is 13.0\arcsec\ at 850\,$\mu$m, before smoothing, with
around 25\% of the total power in the much broader [49\arcsec]
secondary component \citep[see][]{holland13}.

The observations were undertaken whilst moving the telescope
at a constant speed in a so-called {\sc daisy} pattern \citep{holland13},
which provides uniform exposure-time coverage in the central
3\arcmin-diameter region of a field, but useful coverage over
12\arcmin.

Around 10--15\,min was spent integrating on each target, typically
(see Table~\ref{tab:log}), sufficient to detect 850-$\mu$m emission
robustly for $z>4$ far-IR-bright galaxies with a characteristic
temperature of 10--100\,{\sc k}.

The flux-density scale was set using Uranus and Mars, and also
secondary calibrators from the JCMT calibrator list \citep{dempsey13},
with estimated calibration uncertainties amounting to 5\% at
850\,$\mu$m.  Since we visited each target
only once (the handful of exceptions are noted in
Table~\ref{tab:log}), the astrometry of the SCUBA-2 images is expected
to be the same as the JCMT r.m.s.\ pointing accuracy, 2--3\,arcsec.

The data were reduced using the Dynamic Iterative Map-Maker within the
{\sc starlink smurf} package \citep{chapin13} using the `zero-mask'
algorithm, wherein the image is assumed to be free of significant
emission apart from one or more specified regions, in our case a
30-arcsec-diameter circle (larger where appropriate, e.g.\ for
SGP-354388 -- see \S\ref{scuba2fluxes}) centered on the target. This
method is effective at suppressing large-scale noise.  SCUBA-2
observations of flux density calibrators are handled in a similar
manner, generally, so measuring reliable flux densities is
significantly more straightforward than in other situations, as
discussed later in \S\ref{results}.

\subsection{870-$\mu$m continuum imaging with LABOCA}
\label{labocaobs}

Images were also taken with the Large APEX bolometer camera
\citep[LABOCA ---][]{siringo09} mounted on the 12-m Atacama Pathfinder
EXperiment (APEX) telescope\footnote{This publication is based on data
  acquired with APEX, a collaboration between the Max-Planck-Institut
  fur Radioastronomie, the European Southern Observatory, and the
  Onsala Space Observatory.}, on Llano Chajnantor at an altitude of
5,100\,m, in Chile.  LABOCA contains an array of 295 composite
bolometers, arranged as a central channel with nine concentric
hexagons, operating at a central wavelength of 870\,$\mu$m
(806--958\,$\mu$m at half power, so a wider and redder passband than
the SCUBA-2 850-$\mu$m filter) with a {\sc fwhm} resolution of
19.2\arcsec.

All sources were observed using a compact raster pattern in which the
telescope performed a 2.5\arcmin-diameter spiral at constant angular
speed at each of four raster positions, leading to a fully sampled map
over the full 11\arcmin-diameter field of view of LABOCA.  Around
2--4\,hr was spent integrating on each target (see
Table~\ref{tab:log}).  The data were reduced using the BoA software
package, applying standard reductions steps \citep[see e.g.][]{weiss09}.
 
The PWV during the observations was typically between 0.6 and 1.4\,mm,
corresponding to a zenith atmospheric opacity of 0.30--0.55 in the
LABOCA passband.  The flux-density scale was determined to an accuracy
of 10\% using observations of Uranus and Neptune. Pointing was checked
every hour using nearby quasars and was stable. The astrometry of our
LABOCA images, each the result of typically three individual scans,
separated by pointing checks, is expected to be $\sigma
\approx1$--2\arcsec.

\begin{table*}
\caption{Targets and their properties.}
\label{tab:log}
\centering
{\scriptsize
\begin{tabular}{lccr@{\,$\pm$\,}lr@{\,$\pm$\,}lr@{\,$\pm$\,}lr@{\,$\pm$\,}lr@{\,$\pm$\,}lr@{\,$\pm$\,}lc}
  \hline\hline
  \multicolumn{1}{l}{IAU name} &
  \multicolumn{1}{c}{Nickname} &Band&
  \multicolumn{2}{c}{$S_{250}$} &
  \multicolumn{2}{c}{$S_{350}$} &
  \multicolumn{2}{c}{$S_{500}$} &
  \multicolumn{2}{c}{$S_{850}^{\rm peak}$} &
  \multicolumn{2}{c}{$S_{850}^{\rm 45}$} &
  \multicolumn{2}{c}{$S_{850}^{\rm 60}$} &
  \multicolumn{1}{c}{Date} \\
  \multicolumn{1}{l}{} &
  \multicolumn{1}{c}{} &flag&
  \multicolumn{2}{c}{/mJy} &
  \multicolumn{2}{c}{/mJy} &
  \multicolumn{2}{c}{/mJy} &
  \multicolumn{2}{c}{/mJy} &
  \multicolumn{2}{c}{/mJy} &
  \multicolumn{2}{c}{/mJy} &
  \multicolumn{1}{c}{observed$^a$} \\
  \hline
 HATLAS\,085612.1$-$004922 & G09$-$47693 & 1 & 27.4 & 7.3 & 34.4 & 8.1 & 45.4 & 8.6 & 12.5 & 4.0 & 6.4 & 9.1 & 5.4 & 10.8 & 2012$-$04$-$28\\
 HATLAS\,091642.6$+$022147 & G09$-$51190 & 1 & 28.5 & 7.6 & 39.5 & 8.1 & 46.6 & 8.6 & 15.2 & 3.8 & 28.3 & 7.3 & 24.2 & 8.7 & 2012$-$12$-$21\\
 HATLAS\,084113.6$-$004114 & G09$-$59393 & 1 & 24.1 & 7.0 & 43.8 & 8.3 & 46.8 & 8.6 & 23.7 & 3.5 & 27.7 & 5.6 & 12.4 & 9.8 & 2012$-$04$-$27\\
 HATLAS\,090925.0$+$015542 & G09$-$62610 & 1 & 18.6 & 5.4 & 37.3 & 7.4 & 44.3 & 7.8 & 19.5 & 4.9 & 23.1 & 9.0 & 32.7 & 14.4 & 2012$-$03$-$06\\
 HATLAS\,091130.1$-$003846 & G09$-$64889 & 1 & 20.2 & 5.9 & 30.4 & 7.7 & 34.7 & 8.1 & 15.1 & 4.3 & 4.4 & 8.9 & $-$21.2 & 10.0 & 2012$-$12$-$16\\
 HATLAS\,083909.9$+$022718 & G09$-$79552 & 2 & 16.6 & 6.2 & 38.1 & 8.1 & 42.8 & 8.5 & 17.0 & 3.6 & 11.1 & 7.3 & 3.2 & 14.0 & 2013$-$03$-$09\\
 HATLAS\,090419.9$-$013742 & G09$-$79553 & 2 & 14.0 & 5.9 & 36.8 & 8.0 & 35.9 & 8.4 & 16.8 & 3.7 & 20.1 & 7.1 & 14.4 & 10.1 & 2013$-$03$-$09\\
 HATLAS\,084659.0$-$004219 & G09$-$80620 & 2 & 13.5 & 5.0 & 25.3 & 7.4 & 28.4 & 7.7 & 13.2 & 4.3 & 6.8 & 9.8 & $-$9.7 & 9.3 & 2012$-$12$-$16\\
 HATLAS\,085156.0$+$020533 & G09$-$80658 & 2 & 17.8 & 6.4 & 31.6 & 8.3 & 39.5 & 8.8 & 17.6 & 4.1 & 13.6 & 9.4 & 24.0 & 9.4 & 2013$-$03$-$09\\
 HATLAS\,084937.0$+$001455 & G09$-$81106 & 2 & 14.0 & 6.0 & 30.9 & 8.2 & 47.5 & 8.8 & 30.2 & 5.2 & 37.4 & 11.4 & 37.0 & 12.0 & 2012$-$12$-$18\\
 HATLAS\,084059.3$-$000417 & G09$-$81271 & 2 & 15.0 & 6.1 & 30.5 & 8.2 & 42.3 & 8.6 & 29.7 & 3.7 & 35.8 & 6.4 & 44.2 & 10.6 & 2013$-$03$-$09\\
 HATLAS\,090304.2$-$004614 & G09$-$83017 & 2 & 10.2 & 5.7 & 26.4 & 8.0 & 37.2 & 8.8 & 16.1 & 4.4 & 17.9 & 9.4 & 1.7 & 9.1 & 2012$-$12$-$16\\
 HATLAS\,090045.4$+$004125 & G09$-$83808 & 2 & 9.7 & 5.4 & 24.6 & 7.9 & 44.0 & 8.2 & 36.0 & 3.1 & 36.2 & 9.1 & 23.5 & 10.4 & 2012$-$12$-$16\\
 HATLAS\,083522.1$+$005228 & G09$-$84477 & 2 & 20.0 & 6.6 & 27.3 & 8.3 & 31.6 & 9.0 & 7.6 & 3.8 & $-$6.5 & 7.4 & $-$25.8 & 8.9 & 2012$-$04$-$27\\
 HATLAS\,090916.2$+$002523 & G09$-$87123 & 2 & 10.4 & 5.8 & 25.3 & 8.2 & 39.2 & 8.7 & 20.7 & 4.6 & 24.5 & 9.3 & 43.7 & 12.4 & 2012$-$12$-$16\\
 HATLAS\,090855.6$+$015638 & G09$-$100369 & 2 & 15.4 & 5.5 & 17.3 & 7.6 & 32.3 & 8.0 & 13.2 & 3.6 & 22.1 & 8.2 & 14.3 & 9.8 & 2013$-$03$-$09\\
 HATLAS\,090808.9$+$015459 & G09$-$101355 & 3 & 9.5 & 5.5 & 14.6 & 7.9 & 33.4 & 8.3 & 13.5 & 4.9 & $-$2.5 & 10.0 & $-$40.2 & 12.7 & 2012$-$12$-$16\\
 HATLAS\,115415.5$-$010255 & G12$-$34009 & 1 & 30.2 & 7.2 & 36.3 & 8.2 & 60.4 & 8.7 & 39.9 & 4.2 & 38.9 & 9.0 & 38.2 & 17.5 & 2013$-$03$-$09\\
 HATLAS\,114314.6$+$002846 & G12$-$42911 & 1 & 21.2 & 5.8 & 44.1 & 7.4 & 53.9 & 7.7 & 35.4 & 3.6 & 32.8 & 7.0 & 21.0 & 8.0 & 2012$-$04$-$27\\
 HATLAS\,114412.1$+$001812 & G12$-$66356 & 1 & 18.3 & 5.4 & 26.5 & 7.4 & 32.9 & 7.8 & 11.2 & 4.6 & $-$7.5 & 8.8 & $-$2.2 & 12.5 & 2012$-$12$-$18\\
 HATLAS\,114353.5$+$001252 & G12$-$77450 & 2 & 14.8 & 5.1 & 27.3 & 7.4 & 35.9 & 7.7 & 11.9 & 4.1 & $-$0.3 & 7.9 & $-$6.3 & 8.7 & 2012$-$04$-$27\\
 HATLAS\,115012.2$-$011252 & G12$-$78339 & 2 & 17.0 & 6.2 & 30.8 & 8.1 & 31.6 & 9.0 & 18.1 & 4.3 & 31.3 & 8.9 & 33.3 & 11.2 & 2012$-$04$-$27\\
 HATLAS\,115614.2$+$013905 & G12$-$78868 & 2 & 13.1 & 5.9 & 29.5 & 8.2 & 49.0 & 8.5 & 12.2 & 3.5 & 13.6 & 6.4 & 5.8 & 9.6 & 2012$-$04$-$27\\
 HATLAS\,114038.8$-$022811 & G12$-$79192 & 2 & 15.8 & 6.3 & 28.6 & 8.1 & 34.1 & 8.8 & 5.1 & 3.5 & $-$4.3 & 6.4 & $-$17.4 & 7.8 & 2012$-$12$-$21\\
 HATLAS\,113348.0$-$002930 & G12$-$79248 & 2 & 18.4 & 6.2 & 29.5 & 8.2 & 42.0 & 8.9 & 27.6 & 5.0 & 62.4 & 9.8 & 71.3 & 12.0 & 2012$-$12$-$18\\
 HATLAS\,114408.1$-$004312 & G12$-$80302 & 2 & 15.9 & 6.2 & 27.2 & 8.1 & 35.9 & 9.0 & 6.0 & 3.8 & $-$15.0 & 8.9 & $-$28.8 & 9.5 & 2012$-$04$-$27\\
 HATLAS\,115552.7$-$021111 & G12$-$81658 & 2 & 14.9 & 6.1 & 26.5 & 8.1 & 36.8 & 8.7 & 1.0 & 4.4 & $-$25.5 & 8.7 & $-$32.0 & 12.2 & 2012$-$12$-$21\\
 HATLAS\,113331.1$-$003415 & G12$-$85249 & 2 & 13.3 & 6.1 & 25.0 & 8.3 & 31.4 & 8.8 & 4.4 & 2.7 & $-$0.3 & 5.7 & $-$3.3 & 6.6 & 2012$-$12$-$18\\
 HATLAS\,115241.5$-$011258 & G12$-$87169 & 2 & 13.5 & 6.0 & 23.5 & 8.2 & 33.5 & 8.8 & 6.9 & 4.0 & 9.8 & 9.2 & 6.1 & 9.6 & 2012$-$12$-$21\\
 HATLAS\,114350.1$-$005211 & G12$-$87695 & 2 & 19.0 & 6.4 & 23.9 & 8.3 & 30.7 & 8.7 & 15.6 & 3.9 & 2.2 & 7.1 & $-$6.2 & 10.4 & 2012$-$12$-$21\\
 HATLAS\,142208.7$+$001419 & G15$-$21998 & 1 & 36.0 & 7.2 & 56.2 & 8.1 & 62.6 & 8.8 & 13.2 & 3.4 & 7.2 & 7.0 & 7.3 & 9.0 & 2012$-$04$-$26\\
 HATLAS\,144003.9$-$011019 & G15$-$24822 & 1 & 33.9 & 7.1 & 38.6 & 8.2 & 58.0 & 8.8 & 8.0 & 3.5 & 5.8 & 7.5 & 1.4 & 9.0 & 2012$-$04$-$27\\
 HATLAS\,144433.3$+$001639 & G15$-$26675 & 1 & 26.8 & 6.3 & 57.2 & 7.4 & 61.4 & 7.7 & 45.6 & 3.6 & 36.6 & 10.3 & 27.9 & 9.6 & 2012$-$04$-$27\\
 HATLAS\,141250.2$-$000323 & G15$-$47828 & 1 & 28.0 & 7.4 & 35.1 & 8.1 & 45.3 & 8.8 & 19.6 & 4.5 & 15.1 & 9.3 & 10.7 & 10.8 & 2012$-$07$-$28\\
 HATLAS\,142710.6$+$013806 & G15$-$64467 & 1 & 20.2 & 5.8 & 28.0 & 7.5 & 33.4 & 7.8 & 18.7 & 4.9 & 30.7 & 10.8 & 39.2 & 16.2 & 2013$-$03$-$09\\
 HATLAS\,143639.5$-$013305 & G15$-$66874 & 1 & 22.9 & 6.6 & 34.9 & 8.1 & 35.8 & 8.5 & 27.3 & 5.3 & 34.1 & 12.5 & 29.2 & 12.6 & 2012$-$07$-$27\\
 HATLAS\,140916.8$-$014214 & G15$-$82412 & 1 & 21.2 & 6.6 & 30.8 & 8.1 & 41.9 & 8.8 & 17.2 & 4.4 & 9.4 & 8.1 & 6.2 & 10.9 & 2012$-$07$-$28\\
 HATLAS\,145012.7$+$014813 & G15$-$82684 & 2 & 17.3 & 6.4 & 38.5 & 8.1 & 43.2 & 8.8 & 18.5 & 4.1 & 15.3 & 8.2 & 5.5 & 9.3 & 2012$-$04$-$27\\
 HATLAS\,140555.8$-$004450 & G15$-$83543 & 2 & 16.5 & 6.4 & 32.3 & 8.1 & 40.2 & 8.8 & 13.7 & 4.7 & 18.3 & 10.0 & 18.4 & 9.5 & 2012$-$07$-$28\\
 HATLAS\,143522.8$+$012105 & G15$-$83702 & 2 & 14.0 & 6.1 & 30.6 & 8.0 & 33.1 & 8.7 & 7.9 & 4.6 & 4.7 & 8.3 & $-$0.4 & 11.2 & 2012$-$07$-$27\\
 HATLAS\,141909.7$-$001514 & G15$-$84546 & 2 & 11.5 & 4.7 & 23.7 & 7.4 & 30.3 & 7.7 & 19.4 & 5.0 & 10.2 & 9.3 & 7.4 & 12.2 & 2012$-$07$-$27\\
 HATLAS\,142647.8$-$011702 & G15$-$85113 & 2 & 10.5 & 5.7 & 29.6 & 8.2 & 34.9 & 8.7 & 8.7 & 3.4 & 1.6 & 6.9 & 5.2 & 7.5 & 2012$-$04$-$27\\
 HATLAS\,143015.0$+$012248 & G15$-$85592 & 2 & 12.9 & 5.0 & 23.5 & 7.5 & 33.9 & 7.9 & 4.7 & 5.6 & 6.3 & 11.7 & $-$4.3 & 13.7 & 2012$-$07$-$27\\
 HATLAS\,142514.7$+$021758 & G15$-$86652 & 2 & 15.6 & 6.0 & 28.1 & 8.2 & 38.5 & 8.9 & 11.4 & 3.8 & 5.1 & 5.8 & 4.3 & 7.8 & 2012$-$04$-$26\\
 HATLAS\,140609.2$+$000019 & G15$-$93387 & 2 & 15.5 & 6.1 & 23.6 & 8.2 & 35.6 & 8.5 & 8.8 & 3.0 & 14.9 & 6.8 & 15.7 & 8.5 & 2012$-$04$-$27\\
 HATLAS\,144308.3$+$015853 & G15$-$99748 & 2 & 14.0 & 5.8 & 22.4 & 8.3 & 31.5 & 8.8 & 12.2 & 3.8 & 5.0 & 6.4 & 17.9 & 9.7 & 2012$-$04$-$26\\
 HATLAS\,143139.7$-$012511 & G15$-$105504 & 3 & 15.0 & 6.6 & 15.6 & 8.4 & 35.9 & 9.0 & 8.5 & 3.8 & 9.9 & 8.1 & 11.8 & 9.5 & 2012$-$07$-$27\\
 HATLAS\,134040.3$+$323709 & NGP$-$63663 & 1 & 30.6 & 6.8 & 53.5 & 7.8 & 50.1 & 8.1 & 15.5 & 4.1 & 7.9 & 8.3 & $-$12.5 & 9.2 & 2012$-$04$-$28\\
 HATLAS\,131901.6$+$285438 & NGP$-$82853 & 1 & 23.6 & 5.8 & 37.6 & 7.3 & 40.5 & 7.5 & 15.8 & 3.6 & 2.1 & 5.2 & $-$3.8 & 7.8 & 2012$-$06$-$23\\
 HATLAS\,134119.4$+$341346 & NGP$-$101333 & 1 & 32.4 & 7.5 & 46.5 & 8.2 & 52.8 & 9.0 & 24.6 & 3.8 & 17.6 & 8.2 & 13.0 & 9.2 & 2012$-$04$-$28\\
 HATLAS\,125512.4$+$251358 & NGP$-$101432 & 1 & 27.7 & 6.9 & 44.8 & 7.8 & 54.1 & 8.3 & 24.3 & 4.0 & 32.0 & 7.2 & 41.9 & 10.9 & 2012$-$06$-$23\\
 HATLAS\,130823.9$+$254514 & NGP$-$111912 & 1 & 25.2 & 6.5 & 41.5 & 7.6 & 50.2 & 8.0 & 14.9 & 3.9 & 8.8 & 6.7 & 2.3 & 9.1 & 2012$-$04$-$26\\
 HATLAS\,133836.0$+$273247 & NGP$-$113609 & 1 & 29.4 & 7.3 & 50.1 & 8.0 & 63.5 & 8.6 & 21.9 & 3.5 & 12.5 & 6.2 & 9.2 & 9.5 & 2012$-$04$-$26\\
 HATLAS\,133217.4$+$343945 & NGP$-$126191 & 1 & 24.5 & 6.4 & 31.3 & 7.7 & 43.7 & 8.2 & 29.7 & 4.3 & 37.2 & 7.5 & 45.1 & 11.6 & 2012$-$04$-$28\\
 HATLAS\,130329.2$+$232212 & NGP$-$134174 & 1 & 27.6 & 7.3 & 38.3 & 8.4 & 42.9 & 9.4 & 11.4 & 4.0 & 21.3 & 7.4 & 11.7 & 8.9 & 2012$-$04$-$26\\
 HATLAS\,132627.5$+$335633 & NGP$-$136156 & 1 & 29.3 & 7.4 & 41.9 & 8.3 & 57.5 & 9.2 & 23.4 & 3.4 & 29.7 & 4.6 & 27.7 & 9.8 & 2012$-$04$-$26\\
 \hline
\end{tabular}}
\end{table*}

\addtocounter{table}{-1}
\begin{table*}
\caption{Cont...}
{\centering
{\scriptsize
\begin{tabular}{lccr@{\,$\pm$\,}lr@{\,$\pm$\,}lr@{\,$\pm$\,}lr@{\,$\pm$\,}lr@{\,$\pm$\,}lr@{\,$\pm$\,}lc}
  \hline\hline
  \multicolumn{1}{l}{IAU name} &
  \multicolumn{1}{c}{Nickname} &Band&
  \multicolumn{2}{c}{$S_{250}$} &
  \multicolumn{2}{c}{$S_{350}$} &
  \multicolumn{2}{c}{$S_{500}$} &
  \multicolumn{2}{c}{$S_{850}^{\rm peak}$} &
  \multicolumn{2}{c}{$S_{850}^{\rm 45}$} &
  \multicolumn{2}{c}{$S_{850}^{\rm 60}$} &
  \multicolumn{1}{c}{Date} \\
  \multicolumn{1}{l}{} &
  \multicolumn{1}{c}{} &flag&
  \multicolumn{2}{c}{/mJy} &
  \multicolumn{2}{c}{/mJy} &
  \multicolumn{2}{c}{/mJy} &
  \multicolumn{2}{c}{/mJy} &
  \multicolumn{2}{c}{/mJy} &
  \multicolumn{2}{c}{/mJy} &
  \multicolumn{1}{c}{observed$^a$} \\
  \hline
 HATLAS\,J130545.8$+$252953 & NGP$-$136610 & 1 & 23.1 & 6.2 & 39.3 & 7.7 & 46.3 & 8.3 & 19.4 & 3.6 & 34.6 & 7.5 & 29.3 & 9.9 & 2012$-$07$-$12\\
 HATLAS\,J130456.6$+$283711 & NGP$-$158576 & 1 & 23.4 & 6.3 & 38.5 & 7.7 & 38.2 & 8.1 & 13.1 & 4.0 & 12.0 & 7.3 & 15.8 & 10.2 & 2012$-$04$-$26\\
 HATLAS\,J130515.8$+$253057 & NGP$-$168885 & 1 & 21.2 & 6.0 & 35.2 & 7.7 & 45.3 & 8.0 & 26.5 & 3.8 & 17.8 & 7.2 & 4.7 & 8.9 & 2013$-$03$-$09\\
 HATLAS\,J131658.1$+$335457 & NGP$-$172391 & 1 & 25.1 & 7.1 & 39.2 & 8.1 & 52.3 & 9.1 & 15.4 & 3.1 & 7.2 & 6.0 & 5.3 & 8.6 & 2012$-$04$-$26\\
 HATLAS\,J125607.2$+$223046 & NGP$-$185990 & 1 & 24.3 & 7.0 & 35.6 & 8.1 & 41.7 & 8.9 & 33.6 & 4.1 & 18.4 & 9.9 & 13.4 & 12.0 & 2013$-$03$-$09\\
 HATLAS\,J133337.6$+$241541 & NGP$-$190387 & 1 & 25.2 & 7.2 & 41.9 & 8.0 & 63.3 & 8.8 & 37.4 & 3.8 & 33.4 & 8.0 & 29.4 & 10.0 & 2012$-$04$-$26\\
 HATLAS\,J125440.7$+$264925 & NGP$-$206987 & 1 & 24.1 & 7.1 & 39.2 & 8.2 & 50.1 & 8.7 & 22.7 & 3.7 & 17.5 & 6.5 & 25.7 & 9.4 & 2012$-$04$-$26\\
 HATLAS\,J134729.9$+$295630 & NGP$-$239358 & 1 & 21.3 & 6.6 & 28.7 & 8.1 & 33.9 & 8.7 & 15.2 & 5.1 & 39.5 & 13.0 & 61.5 & 15.7 & 2013$-$03$-$09\\
 HATLAS\,J133220.4$+$320308 & NGP$-$242820 & 2 & 18.1 & 6.1 & 35.4 & 7.9 & 33.8 & 8.6 & 14.7 & 3.9 & 10.5 & 7.8 & $-$4.6 & 9.4 & 2012$-$04$-$26\\
 HATLAS\,J130823.8$+$244529 & NGP$-$244709 & 2 & 23.1 & 6.9 & 34.2 & 8.2 & 34.9 & 8.7 & 17.4 & 4.0 & 15.6 & 9.7 & 24.0 & 11.5 & 2013$-$03$-$09\\
 HATLAS\,J134114.2$+$335934 & NGP$-$246114 & 2 & 17.3 & 6.5 & 30.4 & 8.1 & 33.9 & 8.5 & 25.9 & 4.6 & 32.4 & 8.2 & 37.2 & 8.9 & 2012$-$04$-$26\\
 HATLAS\,J131715.3$+$323835 & NGP$-$247012 & 2 & 10.5 & 4.8 & 25.3 & 7.5 & 31.7 & 7.7 & 18.4 & 3.9 & 18.5 & 8.4 & 6.4 & 8.7 & 2013$-$03$-$09\\
 HATLAS\,J131759.9$+$260943 & NGP$-$247691 & 2 & 16.5 & 5.6 & 26.2 & 7.6 & 33.2 & 8.2 & 17.8 & 4.2 & 17.5 & 8.7 & 21.2 & 13.1 & 2013$-$03$-$09\\
 HATLAS\,J133446.1$+$301933 & NGP$-$248307 & 2 & 10.4 & 5.4 & 28.3 & 8.0 & 35.1 & 8.3 & 10.7 & 3.7 & 2.6 & 7.1 & $-$8.5 & 9.1 & 2012$-$04$-$26\\
 HATLAS\,J133919.3$+$245056 & NGP$-$252305 & 2 & 15.3 & 6.1 & 27.7 & 8.1 & 40.0 & 9.4 & 24.0 & 3.5 & 23.5 & 7.6 & 21.2 & 8.7 & 2012$-$04$-$26\\
 HATLAS\,J133356.3$+$271541 & NGP$-$255731 & 2 & 8.4 & 5.0 & 23.6 & 7.7 & 29.5 & 7.9 & 24.6 & 5.2 & 31.0 & 12.4 & 29.5 & 18.4 & 2013$-$03$-$09\\
 HATLAS\,J132731.0$+$334850 & NGP$-$260332 & 2 & 12.2 & 5.8 & 25.1 & 8.1 & 44.4 & 8.6 & 10.1 & 3.2 & 15.9 & 6.0 & 12.0 & 8.8 & 2012$-$04$-$26\\
 HATLAS\,J133251.5$+$332339 & NGP$-$284357 & 2 & 12.6 & 5.3 & 20.4 & 7.8 & 42.4 & 8.3 & 28.9 & 4.3 & 27.4 & 9.9 & 37.0 & 14.4 & 2013$-$03$-$09\\
 HATLAS\,J132419.5$+$343625 & NGP$-$287896 & 2 & 3.4 & 5.7 & 21.8 & 8.1 & 36.4 & 8.7 & 18.7 & 4.3 & $-$8.7 & 8.9 & $-$10.7 & 11.7 & 2013$-$03$-$09\\
 HATLAS\,J131425.9$+$240634 & NGP$-$297140 & 2 & 15.5 & 6.2 & 21.1 & 8.2 & 36.8 & 8.6 & 9.0 & 4.3 & 18.2 & 9.8 & 14.5 & 10.2 & 2013$-$03$-$09\\
 HATLAS\,J132600.0$+$231546 & NGP$-$315918 & 3 & 8.1 & 5.7 & 15.4 & 8.2 & 41.8 & 8.8 & 16.1 & 3.9 & 21.8 & 8.4 & 31.7 & 11.6 & 2013$-$03$-$09\\
 HATLAS\,J132546.1$+$300849 & NGP$-$315920 & 3 & 17.8 & 6.2 & 16.6 & 8.1 & 39.4 & 8.6 & 10.4 & 4.3 & 0.0 & 10.3 & $-$1.5 & 14.2 & 2013$-$03$-$09\\
 HATLAS\,J125433.5$+$222809 & NGP$-$316031 & 3 & 7.0 & 5.5 & 11.4 & 8.2 & 33.2 & 8.6 & 16.8 & 4.0 & 14.1 & 9.3 & 9.1 & 10.9 & 2013$-$03$-$09\\
 HATLAS\,J000124.9$-$354212 & SGP$-$28124 & 1 & 61.6 & 7.7 & 89.1 & 8.3 & 117.7 & 8.8 & 37.2 & 2.6 & 46.7 & 6.0 & 51.6 & 7.8 & 2012$-$12$-$15\\
 HATLAS\,J000124.9$-$354212 & SGP$-$28124$^b$ & 1 & 61.6 & 7.7 & 89.1 & 8.3 & 117.7 & 8.8 & 46.9 & 1.7 & 48.4 & 2.5 & 55.1 & 3.8 & 2013$-$04\\
 HATLAS\,J010740.7$-$282711 & SGP$-$32338 & 2 & 16.0 & 7.1 & 33.2 & 8.0 & 63.7 & 8.7 & 23.1 & 2.9 & 27.9 & 9.4 & 14.3 & 10.0 & 2012$-$12$-$17\\
 HATLAS\,J000018.0$-$333737 & SGP$-$72464 & 1 & 43.4 & 7.6 & 67.0 & 8.0 & 72.6 & 8.9 & 20.0 & 4.2 & 17.2 & 8.9 & 7.5 & 8.2 & 2012$-$12$-$15\\
 HATLAS\,J000624.3$-$323019 & SGP$-$93302 & 1 & 31.2 & 6.7 & 60.7 & 7.7 & 61.7 & 7.8 & 37.1 & 3.7 & 18.4 & 9.1 & 3.6 & 8.3 & 2012$-$12$-$19\\
 HATLAS\,J000624.3$-$323019 & SGP$-$93302$^b$ & 1 & 31.2 & 6.7 & 60.7 & 7.7 & 61.7 & 7.8 & 35.3 & 1.6 & 31.3 & 2.3 & 30.9 & 3.7 & 2013$-$04\\
 HATLAS\,J001526.4$-$353738 & SGP$-$135338 & 1 & 32.9 & 7.3 & 43.6 & 8.1 & 53.3 & 8.8 & 14.7 & 3.8 & 20.8 & 8.0 & 17.9 & 8.4 & 2012$-$12$-$19\\
 HATLAS\,J223835.6$-$312009 & SGP$-$156751 & 1 & 28.4 & 6.9 & 37.7 & 7.9 & 47.6 & 8.4 & 12.6 & 2.0 & 12.0 & 2.9 & 12.5 & 3.5 & 2013$-$04\\
 HATLAS\,J000306.9$-$330248 & SGP$-$196076 & 1 & 28.6 & 7.3 & 28.6 & 8.2 & 46.2 & 8.6 & 32.5 & 4.1 & 32.5 & 9.8 & 32.2 & 11.2 & 2012$-$12$-$15\\
 HATLAS\,J003533.9$-$280302 & SGP$-$208073 & 1 & 28.0 & 7.4 & 33.2 & 8.1 & 44.3 & 8.5 & 19.4 & 2.9 & 19.7 & 4.3 & 18.9 & 6.3 & 2013$-$04\\
 HATLAS\,J001223.5$-$313242 & SGP$-$213813 & 1 & 23.9 & 6.3 & 35.1 & 7.6 & 35.9 & 8.2 & 18.1 & 3.6 & 18.6 & 6.9 & 12.0 & 8.9 & 2012$-$12$-$19\\
 HATLAS\,J001635.8$-$331553 & SGP$-$219197 & 1 & 27.6 & 7.4 & 51.3 & 8.1 & 43.6 & 8.4 & 12.2 & 3.7 & 15.0 & 7.5 & 6.4 & 10.1 & 2012$-$12$-$21\\
 HATLAS\,J002455.5$-$350141 & SGP$-$240731 & 1 & 25.1 & 7.0 & 40.2 & 8.4 & 46.1 & 8.9 & 1.4 & 4.4 & $-$2.7 & 12.2 & $-$7.8 & 10.2 & 2012$-$12$-$21\\
 HATLAS\,J000607.6$-$322639 & SGP$-$261206 & 1 & 22.6 & 6.3 & 45.2 & 8.0 & 59.4 & 8.4 & 45.8 & 3.5 & 56.9 & 8.9 & 65.1 & 12.4 & 2012$-$12$-$18\\
 HATLAS\,J002156.8$-$334611 & SGP$-$304822 & 1 & 23.0 & 6.7 & 40.7 & 8.0 & 41.3 & 8.7 & 19.8 & 3.8 & 38.8 & 8.3 & 35.1 & 9.0 & 2012$-$12$-$21\\
 HATLAS\,J001003.6$-$300720 & SGP$-$310026 & 1 & 23.1 & 6.8 & 33.2 & 8.2 & 42.5 & 8.7 & 10.9 & 3.8 & 17.7 & 7.2 & 13.5 & 8.5 & 2012$-$12$-$15\\
 HATLAS\,J002907.0$-$294045 & SGP$-$312316 & 1 & 20.2 & 6.0 & 29.8 & 7.7 & 37.6 & 8.0 & 10.3 & 3.5 & 19.8 & 7.2 & 10.5 & 8.5 & 2012$-$12$-$19\\
 HATLAS\,J225432.0$-$323904 & SGP$-$317726 & 1 & 20.4 & 6.0 & 35.1 & 7.7 & 39.5 & 8.0 & 19.4 & 3.2 & 7.9 & 5.9 & 10.5 & 7.3 & 2013$-$09$-$01\\
 HATLAS\,J004223.5$-$334340 & SGP$-$354388 & 1 & 26.6 & 8.0 & 39.8 & 8.9 & 53.5 & 9.8 & 40.4 & 2.4 & 46.0 & 5.7 & 57.5 & 7.2 & 2014$-$06$-$30\\
 HATLAS\,J004223.5$-$334340 & SGP$-$354388$^b$ & 1 & 26.6 & 8.0 & 39.8 & 8.9 & 53.5 & 9.8 & 38.7 & 3.2 & 39.9 & 4.7 & 64.1 & 10.9 & 2013$-$10\\
 HATLAS\,J004614.1$-$321826 & SGP$-$380990 & 2 & 14.4 & 5.9 & 45.6 & 8.2 & 40.6 & 8.5 & 7.7 & 1.8 & 6.8 & 2.7 & 7.8 & 3.1 & 2013$-$01\\
 HATLAS\,J000248.8$-$313444 & SGP$-$381615 & 2 & 19.4 & 6.6 & 39.1 & 8.1 & 34.7 & 8.5 & 8.5 & 3.6 & 4.4 & 6.5 & 2.5 & 7.3 & 2012$-$12$-$15\\
 HATLAS\,J223702.2$-$340551 & SGP$-$381637 & 2 & 18.7 & 6.8 & 41.5 & 8.4 & 49.3 & 8.6 & 12.6 & 3.7 & 5.9 & 6.8 & $-$3.1 & 8.3 & 2013$-$09$-$01\\
 HATLAS\,J001022.4$-$320456 & SGP$-$382394 & 2 & 15.7 & 5.9 & 35.6 & 8.1 & 35.9 & 8.6 & 8.0 & 2.4 & 3.5 & 2.9 & 9.1 & 3.9 & 2012$-$09\\
 HATLAS\,J230805.9$-$333600 & SGP$-$383428 & 2 & 16.4 & 5.6 & 32.7 & 7.9 & 35.6 & 8.4 & 8.2 & 2.9 & 4.3 & 4.8 & 7.0 & 6.8 & 2013$-$08$-$19\\
 HATLAS\,J222919.2$-$293731 & SGP$-$385891 & 2 & 13.0 & 8.2 & 45.6 & 9.8 & 59.6 & 11.5 & 20.5 & 3.6 & 21.6 & 7.1 & 11.7 & 10.4 & 2013$-$09$-$01\\
 HATLAS\,J231146.6$-$313518 & SGP$-$386447 & 2 & 10.5 & 6.0 & 33.6 & 8.4 & 34.5 & 8.6 & 22.4 & 3.6 & 34.3 & 8.4 & 29.0 & 11.3 & 2013$-$08$-$19\\
 HATLAS\,J003131.1$-$293122 & SGP$-$392029 & 2 & 18.3 & 6.5 & 30.5 & 8.3 & 35.3 & 8.4 & 13.8 & 3.5 & 17.4 & 6.2 & 20.0 & 8.1 & 2012$-$12$-$19\\
 HATLAS\,J230357.0$-$334506 & SGP$-$424346 & 2 & 0.7 & 5.9 & 25.1 & 8.3 & 31.6 & 8.8 & 10.5 & 3.6 & $-$14.2 & 5.7 & $-$19.1 & 7.6 & 2013$-$08$-$19\\
 HATLAS\,J222737.1$-$333835 & SGP$-$433089 & 2 & 23.8 & 9.4 & 31.5 & 9.7 & 39.5 & 10.6 & 14.8 & 1.7 & 15.6 & 2.9 & 14.7 & 4.1 & 2012$-$09\\
 HATLAS\,J225855.7$-$312405 & SGP$-$499646 & 3 & 5.8 & 5.9 & 10.8 & 8.1 & 41.4 & 8.6 & 18.7 & 3.0 & 15.2 & 5.6 & 11.9 & 6.5 & 2013$-$08$-$19\\
 HATLAS\,J222318.1$-$322204 & SGP$-$499698 & 3 & $-$7.8 & 8.5 & 14.9 & 10.3 & 57.0 & 11.6 & 11.1 & 3.7 & 8.5 & 7.7 & 6.4 & 10.0 & 2013$-$09$-$01\\
 HATLAS\,J013301.9$-$330421 & SGP$-$499828 & 3 & 5.6 & 5.8 & 13.5 & 8.3 & 36.6 & 8.9 & 9.8 & 2.6 & 6.4 & 4.2 & 4.2 & 5.0 & 2013$-$10\\
 \hline
\end{tabular}}}

\noindent
$^a$Targets observed with LABOCA have dates in the format YYYY-MM,
since data were taken over a number of nights.

\noindent
$^b$Targets observed with both LABOCA and SCUBA-2 (previous row).
\end{table*}

\section{Results, analysis and discussion}
\label{results}

In what follows we describe our measurements of 850-$\mu$m
[870\,$\mu$m for LABOCA] flux densities for our candidate ultrared
DSFGs.\footnote{For the handful of objects where data exist from both
  SCUBA-2 and LABOCA, e.g.\ SGP-354388, the measured flux densities
  are consistent.}

\subsection{Measurements of flux density}
\label{scuba2fluxes}

We measured 850- or 870-$\mu$m flux densities via several methods,
each useful in different circumstances, listing the results in
Table~\ref{tab:log}.

In the first method, we searched beam-convolved
images\footnote{Effective beam sizes after convolution: 18.4\arcsec\
  [25.6\arcsec] for the SCUBA-2 850-$\mu$m [LABOCA 870-$\mu$m] data.}
for the brightest peak within a 45-arcsec-diameter circle, centered on
the target coordinates.  For point sources these peaks provide the
best estimates of both flux density and astrometric position. The
accuracy of the latter can only be accurate to a 1\arcsec\ $\times$
1\arcsec\ pixel, but this is better than the expected statistical
accuracy for our generally low-SNR detections, as commonly expressed
by $\sigma_{\rm pos}=0.6\, \theta/{\rm SNR}$, where $\theta$ is the
{\sc fwhm} beam size \citep[see Appendix of][]{ivison07}; it is also
better than the r.m.s.\ pointing accuracy of the telescopes which, at
least for our JCMT imaging, dominates the astrometric budget.  The
uncertainty in the flux density was taken to be the r.m.s.\ noise in a
beam-convolved, 9-arcmin$^2$ box centered on the target, after
rejecting outliers.  We have ignored the small degree of flux boosting
anticipated for a method of this kind, since this is mitigated to a
large degree by the high probability of a single, real submm emitter
being found in the small area we search.

In the second method, we measured flux densities in 45- and
60-arcsec-diameter apertures (the former is shown in the Appendix,
Figs~A\ref{fig:gama09}--\ref{fig:sgp}, where we adopt the same format
used for Fig.~\ref{fig:examples}) using the {\sc aper} routine in
Interactive Data Language \citep[IDL ---][]{landsman93}, following
precisely the recipe outlined by \citet{dempsey13}, with a sky annulus
between 1.5$\times$ and 2.0$\times$ the aperture radius. The apertures
were first centered on the brightest peak within a 45-arcsec-diameter
circle, centered in turn on the target coordinates.  For this method,
the error was measured using 500 aperture/annulus pairs placed at
random across the image.

For the purposes of the redshift determination -- described in the
next section -- we adopted the flux density measured in the
beam-convolved image unless the measurement in a 45-arcsec aperture
was at least $3$-$\sigma^{\rm peak}_{850}$ larger, following the
procedure outlined by \citet{karim13}. For NGP-239358, we adopted the
peak flux density since examination of the image revealed extended
emission that we regard as unreliable; for SGP-354388, we adopted the
60-arcsec aperture measurement because the submm emission is clearly
distributed on that scale (a fact confirmed by our ALMA 3-mm imaging
-- \citealt{oteo16grh}).

We find that 86\% of our sample are detected at $\rm SNR>2.5$ in the
SCUBA-2 and/or LABOCA maps. The median $S_{500}/S_{250}$ color of this
subset falls from 2.15 to 2.08, whilst the median $S_{500}/S_{350}$
color remains at 1.26.  There is no appreciable change in either color
as SNR increases.  We find that 94, 81 and 75\% of the {\sc
  bandflag}=1, 2 and 3 subsets have $\rm SNR>2.5$.  This reflects the
higher reliability of {\sc bandflag}=1 sources, as a result of their
detection in all three SPIRE bands, though the small number (eight) of
sources involved in the {\sc bandflag}=3 subset means the fraction
detected is not determined accurately.

\subsection{Photometric redshifts}
\label{photoz}

Broadly speaking, two approaches have been used to measure the
redshifts of galaxies via the shape of their far-IR/submm SEDs, and to
determine the uncertainty associated with those measurements. One
method uses a library of template SEDs, following \citet{aretxaga03};
the other uses a single template SED, chosen to be representative, as
proposed by \citet{lapi11}, \citet{pearson13} and others.

\begin{figure}
\centerline{\psfig{file=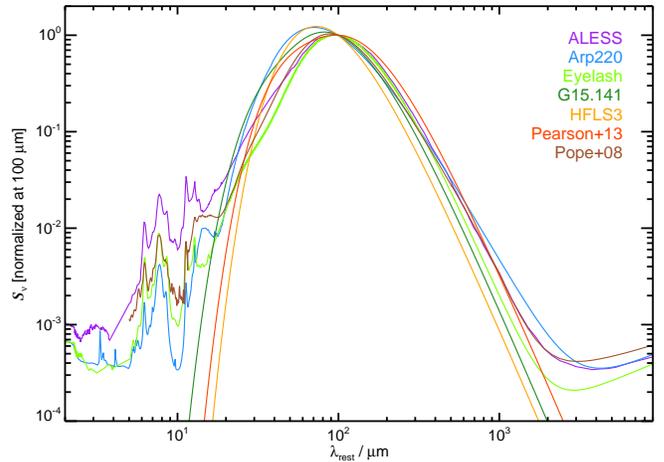,width=3.4in,angle=0}}
\caption{The SED templates used here to determine photometric
  redshifts, normalized in flux density at 100-$\mu$m.  The HFLS3 and
  Arp\,220 SEDs are relatively blue for typical DSFGs, giving us a
  range of plausibly representative templates.}
\label{fig:sedtemplates}
\end{figure}

\begin{figure}
\centerline{\psfig{file=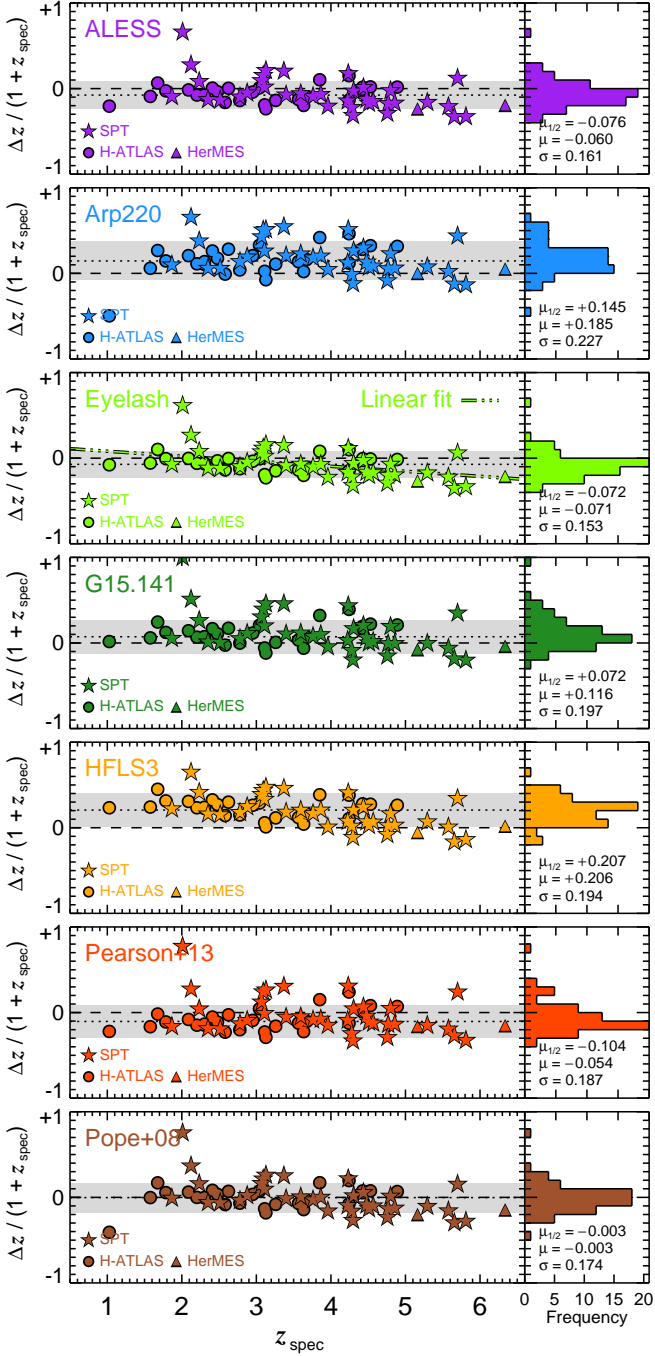,width=3.4in,angle=0}}
\caption{Difference, $(z_{\rm phot} - z_{\rm spec})/(1+z_{\rm spec})$
  or $\Delta z/(1+z_{\rm spec})$, as a function of $z_{\rm spec}$,
  between photometric redshifts determined using the SED templates
  shown in Fig.~\ref{fig:sedtemplates} and the spectroscopic
  redshifts, $z_{\rm spec}$, determined via detections of CO using
  broadband spectrometers for 69 bright DSFGs.  We employed the
  available SPIRE photometric measurements and all additional
  photometry out to 1\,mm, as tabulated by \citet{ivison10fts,
    riechers13, robson14, bussmann13, weiss13, asboth16} and
  \citet{strandet16}.  Approximately the same trend can be seen in
  each panel.  A linear fit of the form $\Delta z/(1+z_{\rm spec})
  \propto -0.059 \times z_{\rm spec}$, which is typical, is shown in
  the Cosmic Eyelash panel.  The statistics noted in each panel
  illustrate the systematic underestimates or overestimates of $z_{\rm
    phot}$ found using the relevant SED templates, and the degree of
  scatter. It is worth noting that the redshifts of the templates are
  recovered accurately, showing that the process works well, e.g.\ in
  the HFLS3 panel, HFLS3 itself can be seen at $z=6.3$ with $\Delta
  z/(1+z_{\rm spec}) =0$. The outlier at $z\sim 2$ is discussed in
  \S\ref{sanitycheck}.  On the basis of these statistics we
  discontinue using the Arp\,220, G15.141, HFLS3 and
  \citeauthor{pearson13} template SEDs in future analyses. }
\label{fig:traditional}
\end{figure}

\begin{figure}
\centerline{\psfig{file=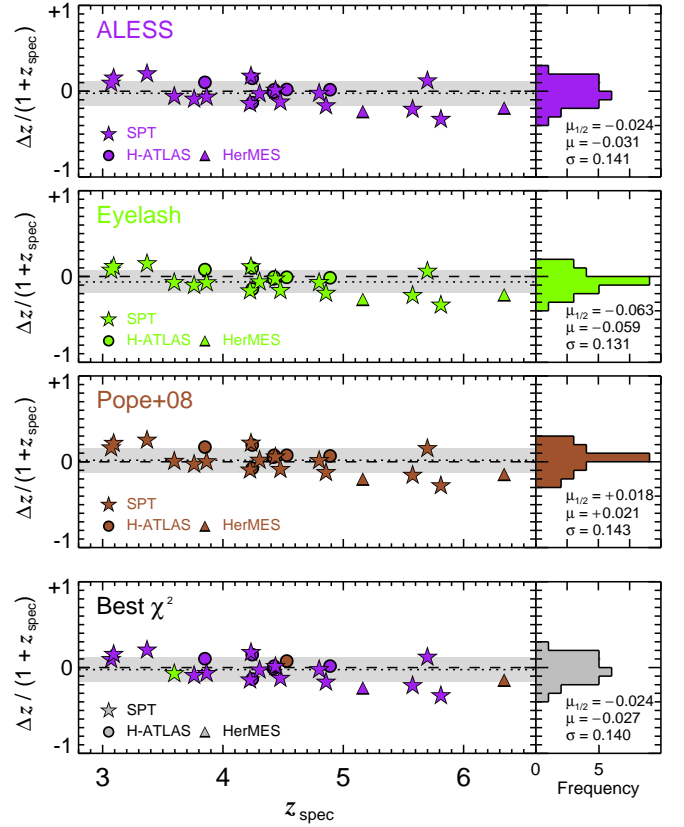,width=3.4in,angle=0}}
\caption{Difference, $\Delta z/(1+z_{\rm spec})$, as a function of
  $z_{\rm spec}$, between photometric redshifts determined using the
  three SEDs shown to be the most effective templates in
  Fig.~\ref{fig:traditional} and the spectroscopic redshifts,
  $z_{\rm spec}$, determined via detections of CO using broadband
  spectrometers for 25 ultrared DSFGs that match the color
  requirements of our sample here, drawn from this paper, from
  \citet{weiss13}, \citet{riechers13}, \citet{asboth16} and
  \citet{strandet16}.  As in Fig.~\ref{fig:traditional}, we employed
  the available SPIRE photometric measurements and all additional
  photometry out to 1\,mm.  The statistics noted in each panel show
  that the systematic underestimates or overestimates of
  $z_{\rm phot}$ found using the relevant SED templates are small, as
  is the scatter.  The lower panel shows $\Delta z/(1+z_{\rm spec})$
  for the template that yields the lowest $\chi^2$ for each ultrared
  DSFG, this being the approach we adopt hereafter to determine the
  redshift distribution of our full sample.  The scatter in this lower
  panel represents the minimum systematic uncertainty in photometric
  redshift since these sources typically have higher S/N photometry
  than our faint, ultrared DSFG candidates.}
\label{fig:sanity}
\end{figure}
 
For the first method, the distribution of measured redshifts and their
associated uncertainties are governed by the choice of template SEDs,
where adopting a broad range of SEDs makes more sense in some
situations than in others. Blindly employing the second method offers
less understanding of the potential systematics and uncertainties.

To characterize the systematics and overall uncertainties, we adopt
seven well-sampled SEDs, all potentially representative of distant
DSFGs: those for HFLS3, Arp\,220, which are both relatively blue for
DSFGs, plus those for the Cosmic Eyelash and G15.141, as well as
synthesized templates from \citet{pope08}, \citet{pearson13} and
\citeauthor{swinbank14} (2014, ALESS) -- see
Fig.~\ref{fig:sedtemplates}.  The \citeauthor{pearson13} template was
synthesised from 40 bright H-ATLAS sources with known
spectroscopic\footnote{It is worth noting a subtle circularity here,
  in that around half of these bright sources were selected as targets
  for broadband spectroscopic observations, e.g.\ with the
  Zpectrometer on the Green Bank Telescope \citep{frayer11, harris12}
  on the basis of rough photometric estimates of their redshifts. The
  resulting bias will be modest, but extreme SEDs may not be fully
  represented.}  redshifts and comprises two modified Planck
functions, $T_{\rm hot}=46.9$\,K and $T_{\rm cold}=23.9$\,K, where the
frequency dependence of the dust emissivity, $\beta$, is set to +2,
and the ratio of cold to hot dust masses is 30.1:1.  The lensed
source, G15.141, is modelled using two greybodies with parameters
taken from \citet{lapi11}, $T_{\rm hot}=60$\,K and $T_{\rm
  cold}=32$\,K, $\beta=+2$ and the ratio of cold to hot dust masses of
50:1.  Fig.~\ref{fig:sedtemplates} shows the diversity of these SEDs
in the rest frame, normalized in flux density at 100\,$\mu$m.

\subsubsection{Training}

Before we use these SED templates to determine the redshifts of our
ultrared DSFGs, we want to estimate any systematic redshift
uncertainties and reject any unsuitable templates, thereby `training'
our technique.  To accomplish this, the SED templates were fitted to
the available photometry for 69 bright DSFGs with SPIRE ($S_{250},
S_{350}, S_{500}$) and $S_{870}$ photometric measurements, the latter
typically from the Submillimeter Array \citep{bussmann13}, and
spectroscopic redshifts determined via detections of CO using
broadband spectrometers \citep[e.g.][]{weiss13, riechers13, asboth16,
  strandet16}. We used accurate filter transmission profiles in each
case, searching for minima in the $\chi^2$ distribution over $0<z_{\rm
  phot}<10$, ignoring possible contamination of the various filters
passbands by bright spectral lines\footnote{With the detection of
  several galaxies in [C\,{\sc ii}] \citep[e.g.][]{oteo16cplus}, we
  are closer to being able to quantify the effect of line emission on
  photometric redshift estimates.} such as [C\,{\sc ii}]
\citep{smail11}.
 
The differences between photometric redshifts estimated in this way
and the measured spectroscopic redshifts for these 69 bright DSFGs
were quantified using the property $(z_{\rm phot} - z_{\rm spec}) / (1
+ z_{\rm spec})$, or $\Delta z/(1+z_{\rm spec})$ hereafter.

Fig.~\ref{fig:traditional} shows the outcome when our seven SED
templates are used to determine photometric redshifts for the 69
bright DSFGs with spectroscopic redshifts.  We might have expected
that the \citeauthor{pearson13} template would yield the most accurate
redshifts for this sample, given that it was synthesized using many of
these same galaxies, but seemingly the inclusion of galaxies with
optical spectroscopic redshifts during its construction has resulted
in a slightly redder SED\footnote{This may be due to blending or
  lensing, or both, where the galaxy with the spectroscopic redshift
  may be just one of a number of contributors to the far-IR flux
  density.}  than the average for those DSFGs with CO spectroscopic
redshifts, resulting in mean and median offsets, $\mu=-0.062$ and
$\mu_{\rm 1/2}=-0.116$, with an r.m.s.\ scatter, $\sigma=0.187$.
While the \citeauthor{pearson13} template fares better than those of
Arp\,220, G15.141 and HFLS3, which have both higher offsets and higher
scatter, as well as a considerable fraction of outliers (defined as
those with $|\Delta z/(1+z_{\rm spec})|>0.3$), at this stage we
discontinued using these four SEDs in the remainder of our
analysis. We retained the three SED templates with $|\mu_{1/2}|<0.1$
and fewer than 10\% outliers for the following important sanity check.

\subsubsection{Sanity check}
\label{sanitycheck}

For this last test we employed the 25 ultrared DSFGs that match the
color requirements\footnote{Although 26 DSFGs meet our color-selection
  criteria, we do not include the extreme outlier, SPT\,0452$-$50,
  which has $\Delta z/(1+z_{\rm spec}) = 0.66, 0.61$ and 0.75 for the
  ALESS, Eyelash and Pope+08 template SEDs, respectively.}  of our
ultrared sample.  Their spectroscopic redshifts have been determined
via detections of CO using broadband spectrometers, typically the 3-mm
receivers at ALMA and NOEMA, drawn partly from the sample in this
paper \citep[see][for the spectroscopic follow-up]{fudamoto16} but
mainly from from the literature \citep{cox11, weiss13, riechers13,
  asboth16, strandet16}.

Without altering our redshift-fitting procedure, we employed the
available SPIRE photometric measurements together with all additional
photometry out to 1\,mm.  For each source we noted the redshift and
the template with the best $\chi^2$.  Fig.~\ref{fig:sanity} shows
$\Delta z/(1+z_{\rm spec})$ as a function of $z_{\rm spec}$ and we can
see that the Cosmic Eyelash and the synthesized templates from
\citeauthor{swinbank14} (ALESS) and \citeauthor{pope08} have excellent
predictive capabilities, with $|\mu_{1/2}|\ls0.06$ and $\sigma\sim
0.14$.

The lower panel of Fig.~\ref{fig:sanity} shows
$\Delta z/(1+z_{\rm spec})$ versus $z_{\rm spec}$ for the SED template
that yields the best $\chi^2$ for each ultrared DSFG, where
$\mu_{1/2}=-0.024$.  The scatter seen in this plot is representative
of the minimum systematic uncertainty in determining photometric
redshifts for ultrared galaxies, $\sigma\sim 0.14$, given that the
photometry for these brighter sources tends to be of a relatively high
quality.  Despite a marginally higher scatter than the best individual
SED templates, we adopt the photometric redshifts with the lowest
$\chi^2$ values hereafter. 

\subsubsection{The effect of the CMB}

We have quantified the well-known effect of the CMB on the SED shape
\citep{dacunha13,zhang16} by using a dual-greybody 30\,{\sc k} +
60\,{\sc k} parameterization of the Cosmic Eyelash -- the prescription
of \citet{ivison10fts}. At $z=2.3$, the Cosmic Eyelash is affected
negligably by the CMB effect -- of the two greybodies, the coolest is
affected most, and it changes by just $\sim4$\,mK compared with $z=0$.
We therefore ignore this and modify the parameterized $z=2.3$ SED to
account for the effect of the CMB at progressively higher redshifts,
then fit the unmodified Cosmic Eyelash SED to monochromatic flux
densities drawn from these modified SEDs at $\lambda_{\rm obs}=250$,
350, 500 and 870\,$\mu$m.  The CMB effect causes us to underestimate
$(1+z)$ by 0.03, 0.05, 0.10 and 0.18 at $z=4$, 6, 8 and 10. Thus, the
effect is small, even at the highest plausible redshifts; moreover,
since the effect biases our redshifts to lower values, our estimate of
the space density of ultrared DSFGs at $z>4$ presented in
\S\ref{spacedensity} will be biased lower rather than higher.

\subsubsection{Redshift trends} 

As an aside, a trend --- approximately the same trend in each case ---
can be seen in each panel of
Figs~\ref{fig:traditional}--\ref{fig:sanity}, with $\Delta z/(1+z_{\rm
  spec})$ decreasing numerically with increasing redshift.  The
relationship takes the form $\Delta z/(1+z_{\rm spec}) \propto
-0.059_{-0.014}^{+0.016} \times z_{\rm spec}$ for the Cosmic Eyelash,
and a consistent trend is seen for the other SED templates.  Were we
to correct for this trend, the typical scatter in $\Delta z/(1+z_{\rm
  spec})$ would fall to $\sim 0.10$. This effect is much stronger than
can be ascribed to the influence of the CMB and betrays a link between
redshift and \Td, which in turn may be related to the relationship
between redshift and \lir\ seen by \citet{sym13}, though disentangling
the complex relationships between \Td, \Md, \lir, starburst size and
redshift is extraordinarily challenging, even if the cross-section to
gravitational lensing were constant with distance, which it is not
(see \S\ref{spacedensity}).  By considering a greybody at the
temperature of each of the templates in our library, we can deduce
that an offset between the photometric and spectroscopic redshifts
corresponds to a change in dust temperature of

\begin{equation}
\Delta T_{\rm dust} = T_{\rm dust}^{\rm grey} \left(
  \frac{1+z_{\rm phot}}{1+z{_{\rm spec}}} - 1 \right) ,
\end{equation}

\noindent
where $\Delta T_{\rm dust}$ is difference between the dust temperature
of the source and the temperature of the template SED, $T_{\rm
  dust}^{\rm SED}$. Using the offset between the photometric and
spectroscopic redshifts for the Cosmic Eyelash template, we estimate
that the typical dust temperature of the sources in our sample becomes
warmer on average by $9.4_{-3.3}^{+4.8}$\,{\sc k} as we move from
$z=2$ ($-0.7$\,{\sc k}) to $z=6$ ($+8.7$\,{\sc k}). We find consistent
results for the \citeauthor{pope08} and ALESS template SEDs, where
$\Delta T_{\rm dust}=7.5_{-3.1}^{+4.0}$\,{\sc k} and
$7.1_{-3.0}^{+3.9}$\,{\sc k}, respectively. We do not reproduce the
drop of 10\,{\sc k} between low and high redshift reported by
\citet{sym13} --- quite the reverse, in fact. This may be related to
the higher fraction of gravitationally lensed (and thus intrisically
less luminous) galaxies expected in the bright sample we have used
here to calibrate and test our photometric redshift technique
(\S\ref{spacedensity}). As with the CMB effect, the observed evolution
in temperature with redshift predominantly biases our photometric
redshifts to lower values, re-inforcing the conservative nature of our
estimate of the space density of ultrared DSFGs at $z>4$.

It is also worth noting that the correlation between \lir\ and
redshift -- discussed later in \S\ref{nzsummary} and probably due in
part to the higher flux density limits at $z>5$ -- may mean that
optical depth effects become more influential at the highest
redshifts, with consequences for the evolution of DSFG SEDs that are
difficult to predict.

\subsubsection{Ultimate test of $z_{\rm phot}$ reliability}

Finally, we employed the refined SED fitting procedure outlined above
to determine the redshift distribution of our full sample of ultrared
DSFGs.

\begin{figure}
\centerline{\psfig{file=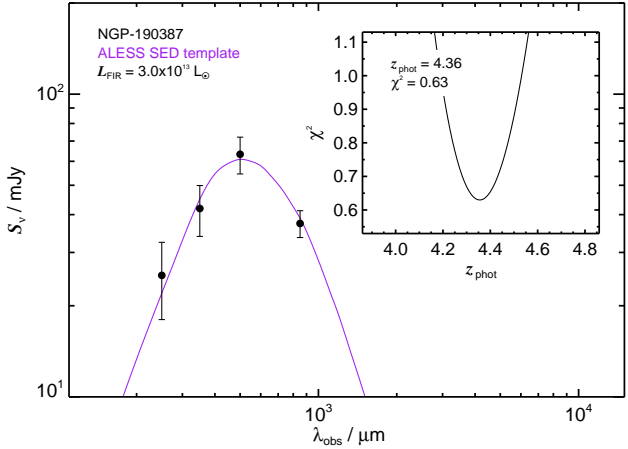,width=3.3in,angle=0}}
\caption{SPIRE and SCUBA-2 photometry for one of our ultrared galaxies
  with a spectroscopic redshift, $z_{\rm spec}=4.42$
  \citep{fudamoto16}, and the best fit to the data,
  $z=4.36^{+0.37}_{-0.26}$ in this case made using the ALESS SED
  template of \citet{swinbank14}.}
\label{fig:sedfitting}
\end{figure}

\begin{figure}
\centerline{\psfig{file=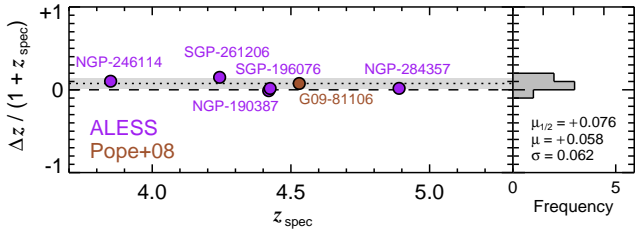,width=3.3in,angle=0}}
\caption{Predictive power of our photometric redshifts, as judged
  using the six ultrared galaxies with spectroscopic redshifts from
  our sample \citep{fudamoto16}, on the same scale used earlier in
  Figs~\ref{fig:traditional} and \ref{fig:sanity}.}
\label{fig:finaltest}
\end{figure}

As a final test of $z_{\rm phot}$ reliability,
Fig.~\ref{fig:sedfitting} shows the best-fitting photometric redshift
for one of the sources, NGP$-$190387, for which we have secured a
spectroscopic redshift using ALMA or NOEMA \citep{fudamoto16} and in
Fig.~\ref{fig:finaltest} we present the photometric redshifts of all
of the six ultrared DSFGs for which we have determined secure
spectroscopic redshifts.

We find $|\mu_{1/2}|=+0.08$ and $\sigma=0.06$, and the r.m.s.\ scatter
around $\Delta z/(1+z_{\rm spec}) =0$ is 0.08, consistent with
expectations\footnote{An appropriate comparison because the scatter
  induced by the $\Delta z/(1+z_{\rm spec}) \propto -0.059\times z$
  trend across $z=3.8$--4.9 will be small.} set by the scatter ($\sim
0.10$) seen earlier amongst the trend-corrected redshifts determined
using the Cosmic Eyelash SED.

\begin{figure}
\centerline{\psfig{file=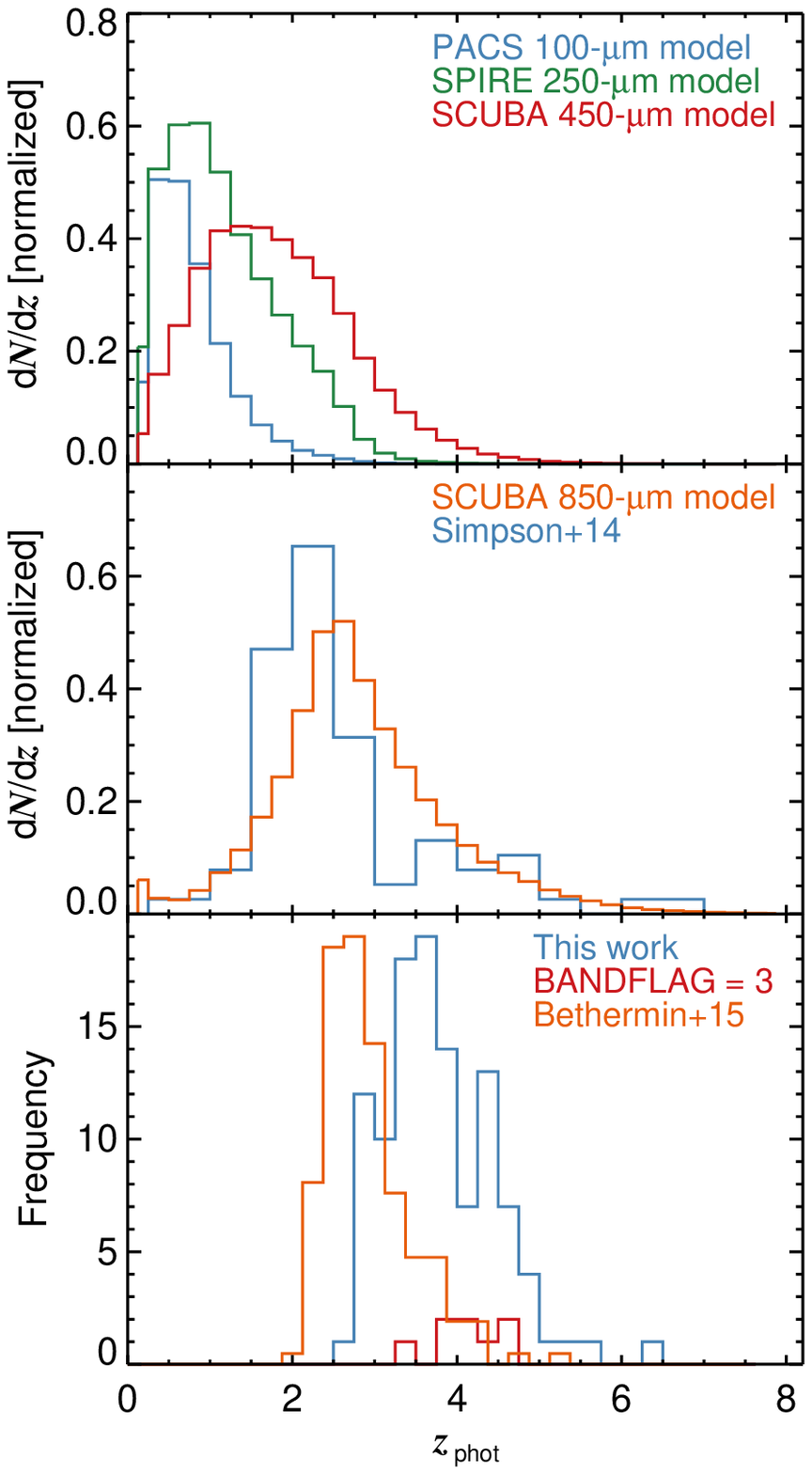,width=2.6in,angle=0}} 
\caption{Redshift histograms from \citet{bethermin15}, representing a
  phenomenological model of galaxy evolution \citep{bethermin12}, with
  the expected redshift distributions for PACS at 100\,$\mu$m
  ($S_{100}>9$\,mJy), SPIRE at 250\,$\mu$m ($S_{250}>20$\,mJy),
  SCUBA-2 at 450\,$\mu$m ($S_{450}>5$\,mJy) in the upper panel. In the
  middle panel we show the \citeauthor{bethermin15} redshift
  distribution predicted for SCUBA-2 at 850\,$\mu$m
  ($S_{850}>4$\,mJy), alongside the redshifts measured for the LABOCA
  870-$\mu$m-selected LESS sample ($S_{870}>3.5$\,mJy) by
  \citet{simpson14}.  In the lower panel we show the histogram of
  redshifts for our sample of ultrared galaxies, where for each galaxy
  we have adopted the redshift corresponding to the best $\chi^2$ fit,
  found with the SED templates used in Fig.~\ref{fig:sanity}.  The
  subset (of eight galaxies) with {\sc bandflag}=3, i.e.\ those
  selected from 500-$\mu$m residual maps, are shown in red.  Our
  ultrared DSFGs typically lie $\delta z\approx 1.5$ redward of the
  870-$\mu$m-selected sample.  Comparison of the observed photometric
  redshift distribution for our ultrared DSFGs with that expected by
  the \citet{bethermin15} model (for sources selected with our flux
  limits and color criteria) reveals a significant mismatch.}
\label{fig:finalhist}
\end{figure}

\subsubsection{Summary of $z_{\rm phot}$ and \lir\ statistics}
\label{nzsummary}

In Table~\ref{tab:photoz}, we list the photometric redshifts (and
luminosities, measured in the rest-frame across 8--1000\,$\mu$m) for
each source in our sample, with uncertainties determined from a Monte
Carlo treatment of the observed flux densities and their respective
uncertainties.

\begin{table}
    \caption{Targets and their photometric redshift properties.}
    \label{tab:photoz}
    \centering
    {\scriptsize
\begin{tabular}{l l r l l r}
    \hline\hline
    \multicolumn{1}{l}{Nickname} &
    \multicolumn{1}{c}{$z$} &
    \multicolumn{1}{c}{$\log_{10}\left(L_{\textrm{FIR}}\right)$} &
    \multicolumn{1}{l}{Nickname} &
    \multicolumn{1}{c}{$z$} &
    \multicolumn{1}{c}{$\log_{10}\left(L_{\textrm{FIR}}\right)$} \\
    \hline
G09-47693 & $3.12^{+0.39}_{-0.33}$ & $13.01^{+0.14}_{-0.07}$ &
NGP-136610 & $4.27^{+0.51}_{-0.51}$ & $13.40^{+0.09}_{-0.12}$\\
G09-51190 & $3.83^{+0.58}_{-0.48}$ & $13.31^{+0.11}_{-0.12}$ &
NGP-158576 & $3.15^{+0.36}_{-0.29}$ & $13.00^{+0.12}_{-0.07}$\\
G09-59393 & $3.70^{+0.35}_{-0.26}$ & $13.28^{+0.05}_{-0.09}$ &
NGP-168885 & $4.09^{+0.42}_{-0.30}$ & $13.32^{+0.06}_{-0.08}$\\
G09-62610 & $3.70^{+0.44}_{-0.26}$ & $13.15^{+0.13}_{-0.06}$ &
NGP-172391 & $3.27^{+0.34}_{-0.26}$ & $13.08^{+0.09}_{-0.06}$\\
G09-64889 & $3.48^{+0.48}_{-0.40}$ & $13.10^{+0.09}_{-0.14}$ &
NGP-185990 & $4.47^{+0.49}_{-0.37}$ & $13.42^{+0.06}_{-0.06}$\\
G09-79552 & $3.59^{+0.34}_{-0.26}$ & $13.11^{+0.09}_{-0.06}$ &
NGP-190387 & $4.36^{+0.37}_{-0.26}$ & $13.49^{+0.05}_{-0.06}$\\
G09-79553 & $3.66^{+0.39}_{-0.30}$ & $13.08^{+0.11}_{-0.07}$ &
NGP-206987 & $4.07^{+0.06}_{-0.60}$ & $13.31^{+0.02}_{-0.13}$\\
G09-80620 & $4.01^{+0.22}_{-0.78}$ & $13.07^{+0.06}_{-0.19}$ &
NGP-239358 & $3.47^{+0.52}_{-0.49}$ & $13.09^{+0.10}_{-0.15}$\\
G09-80658 & $4.07^{+0.09}_{-0.72}$ & $13.20^{+0.03}_{-0.17}$ &
NGP-242820 & $3.41^{+0.44}_{-0.30}$ & $13.02^{+0.13}_{-0.06}$\\
G09-81106 & $4.95^{+0.13}_{-0.73}$ & $13.43^{+0.04}_{-0.13}$ &
NGP-244709 & $3.48^{+0.42}_{-0.40}$ & $13.14^{+0.07}_{-0.12}$\\
G09-81271 & $4.62^{+0.46}_{-0.38}$ & $13.39^{+0.05}_{-0.09}$ &
NGP-246114 & $4.35^{+0.51}_{-0.46}$ & $13.30^{+0.08}_{-0.10}$\\
G09-83017 & $3.99^{+0.53}_{-0.34}$ & $13.09^{+0.12}_{-0.08}$ &
NGP-247012 & $4.59^{+0.16}_{-0.71}$ & $13.21^{+0.04}_{-0.16}$\\
G09-83808 & $5.66^{+0.06}_{-0.76}$ & $13.51^{+0.02}_{-0.11}$ &
NGP-247691 & $3.90^{+0.51}_{-0.45}$ & $13.15^{+0.08}_{-0.13}$\\
G09-84477 & $2.94^{+0.44}_{-0.39}$ & $12.83^{+0.15}_{-0.09}$ &
NGP-248307 & $3.59^{+0.36}_{-0.36}$ & $12.96^{+0.10}_{-0.10}$\\
G09-87123 & $4.28^{+0.52}_{-0.34}$ & $13.17^{+0.12}_{-0.06}$ &
NGP-252305 & $4.34^{+0.43}_{-0.38}$ & $13.29^{+0.06}_{-0.09}$\\
G09-100369 & $3.79^{+0.61}_{-0.46}$ & $13.05^{+0.09}_{-0.13}$ &
NGP-255731 & $4.94^{+0.73}_{-0.66}$ & $13.30^{+0.09}_{-0.15}$\\
G09-101355 & $4.20^{+0.70}_{-0.39}$ & $13.03^{+0.16}_{-0.08}$ &
NGP-260332 & $3.50^{+0.38}_{-0.29}$ & $12.96^{+0.10}_{-0.08}$\\
G12-34009 & $4.53^{+0.37}_{-0.31}$ & $13.51^{+0.05}_{-0.06}$ &
NGP-284357 & $4.99^{+0.44}_{-0.45}$ & $13.40^{+0.05}_{-0.10}$\\
G12-42911 & $4.33^{+0.31}_{-0.26}$ & $13.45^{+0.05}_{-0.07}$ &
NGP-287896 & $4.54^{+0.53}_{-0.37}$ & $13.15^{+0.10}_{-0.09}$\\
G12-66356 & $3.66^{+0.19}_{-0.72}$ & $13.04^{+0.06}_{-0.19}$ &
NGP-297140 & $3.41^{+0.57}_{-0.44}$ & $12.91^{+0.15}_{-0.11}$\\
G12-77450 & $3.53^{+0.46}_{-0.31}$ & $12.99^{+0.14}_{-0.07}$ &
NGP-315918 & $4.32^{+0.54}_{-0.33}$ & $13.10^{+0.11}_{-0.07}$\\
G12-78339 & $4.41^{+0.98}_{-0.70}$ & $13.31^{+0.17}_{-0.18}$ &
NGP-315920 & $3.88^{+0.33}_{-0.89}$ & $13.05^{+0.07}_{-0.21}$\\
G12-78868 & $3.58^{+0.34}_{-0.26}$ & $13.04^{+0.08}_{-0.08}$ &
NGP-316031 & $4.65^{+0.68}_{-0.47}$ & $13.10^{+0.13}_{-0.07}$\\
G12-79192 & $2.95^{+0.38}_{-0.36}$ & $12.80^{+0.12}_{-0.12}$ &
SGP-28124 & $3.93^{+0.08}_{-0.45}$ & $13.65^{+0.02}_{-0.09}$\\
G12-79248 & $6.43^{+0.81}_{-0.89}$ & $13.76^{+0.11}_{-0.14}$ &
SGP-28124* & $3.80^{+0.02}_{-0.42}$ & $13.61^{+0.00}_{-0.11}$\\
G12-80302 & $3.06^{+0.39}_{-0.35}$ & $12.83^{+0.12}_{-0.10}$ &
SGP-72464 & $3.06^{+0.21}_{-0.19}$ & $13.23^{+0.07}_{-0.05}$\\
G12-81658 & $2.93^{+0.38}_{-0.42}$ & $12.77^{+0.12}_{-0.14}$ &
SGP-93302 & $3.91^{+0.27}_{-0.22}$ & $13.46^{+0.04}_{-0.07}$\\
G12-85249 & $2.87^{+0.37}_{-0.36}$ & $12.70^{+0.11}_{-0.12}$ &
SGP-93302* & $3.79^{+0.24}_{-0.21}$ & $13.43^{+0.04}_{-0.07}$\\
G12-87169 & $3.26^{+0.51}_{-0.39}$ & $12.85^{+0.13}_{-0.12}$ &
SGP-135338 & $3.06^{+0.33}_{-0.26}$ & $13.08^{+0.11}_{-0.04}$\\
G12-87695 & $3.68^{+0.58}_{-0.53}$ & $13.09^{+0.09}_{-0.14}$ &
SGP-156751 & $2.93^{+0.24}_{-0.22}$ & $12.97^{+0.08}_{-0.04}$\\
G15-21998 & $2.91^{+0.20}_{-0.19}$ & $13.10^{+0.06}_{-0.05}$ &
SGP-196076 & $4.51^{+0.47}_{-0.39}$ & $13.42^{+0.07}_{-0.06}$\\
G15-24822 & $2.77^{+0.27}_{-0.27}$ & $12.97^{+0.09}_{-0.08}$ &
SGP-208073 & $3.48^{+0.40}_{-0.28}$ & $13.18^{+0.06}_{-0.08}$\\
G15-26675 & $4.36^{+0.25}_{-0.21}$ & $13.55^{+0.04}_{-0.05}$ &
SGP-213813 & $3.49^{+0.40}_{-0.32}$ & $13.15^{+0.07}_{-0.10}$\\
G15-47828 & $3.52^{+0.50}_{-0.39}$ & $13.20^{+0.09}_{-0.11}$ &
SGP-219197 & $2.94^{+0.25}_{-0.24}$ & $13.03^{+0.08}_{-0.07}$\\
G15-64467 & $3.75^{+0.55}_{-0.49}$ & $13.15^{+0.09}_{-0.14}$ &
SGP-240731 & $2.70^{+0.27}_{-0.25}$ & $12.88^{+0.10}_{-0.09}$\\
G15-66874 & $4.07^{+0.57}_{-0.49}$ & $13.30^{+0.10}_{-0.11}$ &
SGP-261206 & $5.03^{+0.58}_{-0.47}$ & $13.64^{+0.09}_{-0.10}$\\
G15-82412 & $3.96^{+0.15}_{-0.70}$ & $13.20^{+0.04}_{-0.16}$ &
SGP-304822 & $4.33^{+0.63}_{-0.51}$ & $13.41^{+0.12}_{-0.12}$\\
G15-82684 & $3.65^{+0.38}_{-0.25}$ & $13.13^{+0.11}_{-0.06}$ &
SGP-310026 & $3.12^{+0.38}_{-0.31}$ & $12.97^{+0.12}_{-0.07}$\\
G15-83543 & $3.53^{+0.42}_{-0.34}$ & $13.05^{+0.12}_{-0.09}$ &
SGP-312316 & $3.17^{+0.41}_{-0.32}$ & $12.94^{+0.12}_{-0.08}$\\
G15-83702 & $3.27^{+0.39}_{-0.36}$ & $12.90^{+0.12}_{-0.12}$ &
SGP-317726 & $3.69^{+0.39}_{-0.30}$ & $13.20^{+0.06}_{-0.10}$\\
G15-84546 & $4.34^{+0.56}_{-0.53}$ & $13.19^{+0.10}_{-0.14}$ &
SGP-354388 & $5.35^{+0.56}_{-0.52}$ & $13.68^{+0.08}_{-0.08}$\\
G15-85113 & $3.40^{+0.37}_{-0.34}$ & $12.90^{+0.09}_{-0.11}$ &
SGP-354388* & $5.43^{+0.84}_{-0.72}$ & $13.69^{+0.12}_{-0.13}$\\
G15-85592 & $3.39^{+0.49}_{-0.39}$ & $12.89^{+0.15}_{-0.13}$ &
SGP-32338 & $3.93^{+0.26}_{-0.24}$ & $13.24^{+0.05}_{-0.04}$\\
G15-86652 & $3.43^{+0.44}_{-0.35}$ & $12.97^{+0.11}_{-0.09}$ &
SGP-380990 & $2.84^{+0.22}_{-0.21}$ & $12.84^{+0.06}_{-0.07}$\\
G15-93387 & $3.24^{+0.50}_{-0.33}$ & $12.87^{+0.12}_{-0.08}$ &
SGP-381615 & $2.98^{+0.29}_{-0.29}$ & $12.91^{+0.09}_{-0.09}$\\
G15-99748 & $3.98^{+0.25}_{-0.79}$ & $13.06^{+0.05}_{-0.20}$ &
SGP-381637 & $3.30^{+0.28}_{-0.25}$ & $13.06^{+0.08}_{-0.07}$\\
G15-105504 & $3.43^{+0.64}_{-0.53}$ & $12.87^{+0.16}_{-0.13}$ &
SGP-382394 & $2.96^{+0.29}_{-0.26}$ & $12.84^{+0.08}_{-0.08}$\\
NGP-63663 & $3.08^{+0.23}_{-0.22}$ & $13.11^{+0.08}_{-0.06}$ &
SGP-383428 & $3.08^{+0.33}_{-0.30}$ & $12.88^{+0.10}_{-0.09}$\\
NGP-82853 & $3.66^{+0.06}_{-0.61}$ & $13.17^{+0.02}_{-0.15}$ &
SGP-385891 & $3.70^{+0.29}_{-0.24}$ & $13.20^{+0.07}_{-0.06}$\\
NGP-101333 & $3.53^{+0.34}_{-0.27}$ & $13.30^{+0.06}_{-0.09}$ &
SGP-386447 & $4.89^{+0.78}_{-0.73}$ & $13.41^{+0.13}_{-0.17}$\\
NGP-101432 & $3.65^{+0.36}_{-0.28}$ & $13.31^{+0.05}_{-0.10}$ &
SGP-392029 & $3.42^{+0.47}_{-0.32}$ & $13.00^{+0.13}_{-0.06}$\\
NGP-111912 & $3.27^{+0.36}_{-0.26}$ & $13.09^{+0.10}_{-0.06}$ &
SGP-424346 & $3.99^{+0.45}_{-0.39}$ & $12.95^{+0.10}_{-0.10}$\\
NGP-113609 & $3.43^{+0.34}_{-0.20}$ & $13.22^{+0.09}_{-0.04}$ &
SGP-433089 & $3.60^{+0.08}_{-0.62}$ & $13.11^{+0.01}_{-0.13}$\\
NGP-126191 & $4.33^{+0.45}_{-0.46}$ & $13.37^{+0.07}_{-0.08}$ &
SGP-499646 & $4.68^{+0.49}_{-0.34}$ & $13.14^{+0.10}_{-0.05}$\\
NGP-134174 & $2.98^{+0.34}_{-0.31}$ & $12.98^{+0.12}_{-0.07}$ &
SGP-499698 & $4.22^{+0.39}_{-0.38}$ & $13.00^{+0.09}_{-0.11}$\\
NGP-136156 & $3.95^{+0.06}_{-0.57}$ & $13.33^{+0.01}_{-0.12}$ &
SGP-499828 & $3.88^{+0.49}_{-0.41}$ & $12.88^{+0.10}_{-0.09}$\\
\hline
\end{tabular}}
\end{table}

We present a histogram of photometric redshifts for our sample of
ultrared galaxies in Fig.~\ref{fig:finalhist}, where for each galaxy
we have adopted the redshift corresponding to the best $\chi^2$ fit,
found with the SED templates used in Fig.~\ref{fig:sanity}.  In the
upper panels of Fig.~\ref{fig:finalhist} we show redshift histograms
from \citet{bethermin15}, representing a phenomenological model of
galaxy evolution \citep{bethermin12}, with the expected redshift
distributions for PACS at 100\,$\mu$m ($S_{100}>9$\,mJy), SPIRE at
250\,$\mu$m ($S_{250}>20$\,mJy), SCUBA-2 at 450\,$\mu$m
($S_{450}>5$\,mJy), SCUBA-2 at 850\,$\mu$m ($S_{850}>4$\,mJy), cf.\
the redshifts measured for the LABOCA 870-$\mu$m-selected LESS sample
($S_{870}>3.5$\,mJy) by \citet{simpson14}.

Our {\it Herschel}-selected ultrared galaxies span $2.7<z_{\rm
  phot}<6.4$, and typically lie $\delta z\approx 1.5$ redward of the
870-$\mu$m-selected sample, showing that our technique can be usefully
employed to select intense, dust-enshrouded starbursts at the highest
redshifts. We find that $33\pm 6$\% of our full sample
\citep[1-$\sigma$ errors,][]{gehrels86} and $63_{-24}^{+20}$\% of our
{\sc bandflag}=3 subset (see overlaid red histogram in lower panel of
Fig.~\ref{fig:finalhist}) lie at $z_{\rm phot}>4$.  In an ultrared
sample comprised largely of faint 500-$\mu$m risers, we find a median
value of $\hat{z}_{\rm phot}=3.66$, a mean of 3.79 and an
interquartile range, 3.30--4.27. This supports the relation between
the SED peak and redshift observed by \citet{swinbank14}, who found
median redshifts of $2.3\pm0.2$, $2.5\pm0.3$ and $3.5\pm0.5$ for
870-$\mu$m-selected DSFGs with SEDs peaking at 250, 350 and
500\,$\mu$m.

\begin{figure}
\centerline{\psfig{file=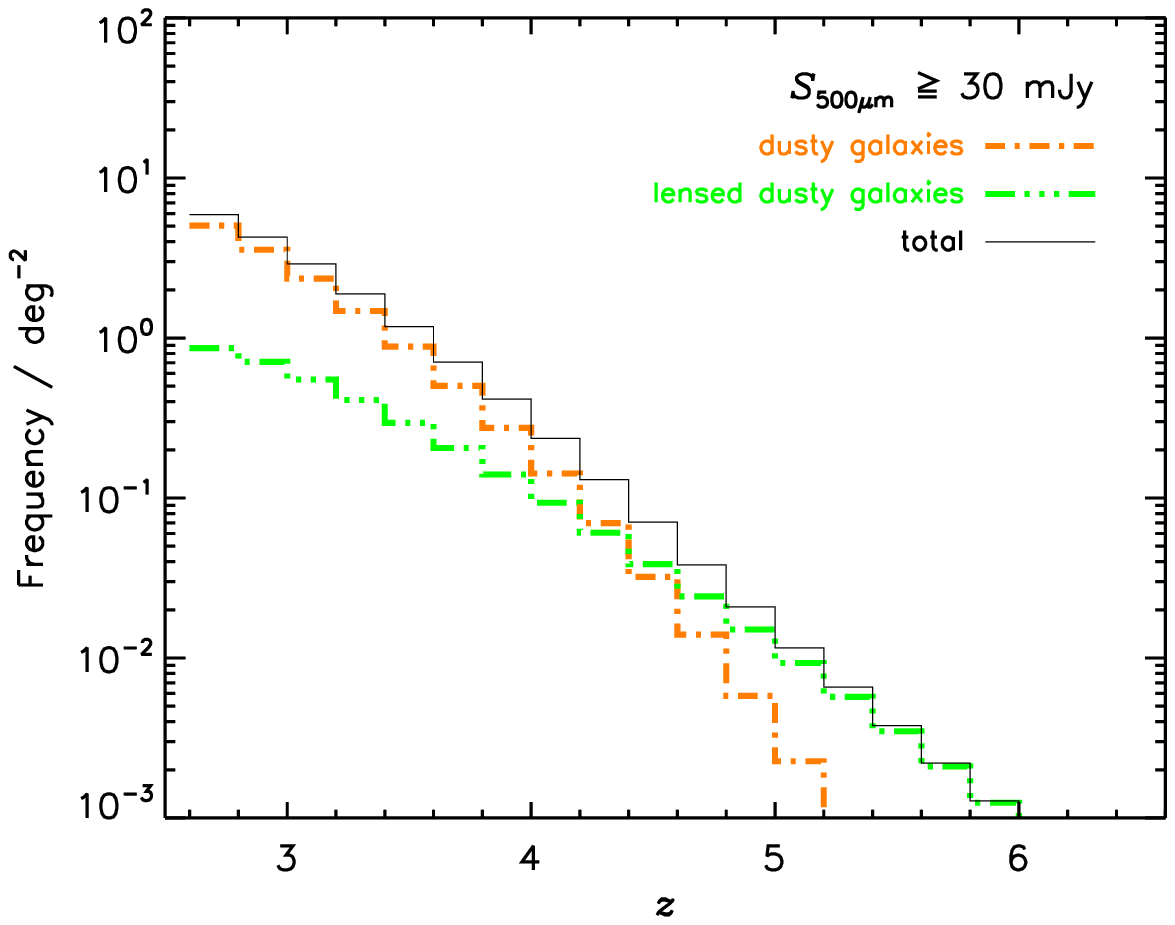,width=3.4in,angle=0}}
\caption{Redshift distribution of $S_{500}>30$-mJy sources from the
  physical model of \citet{cai13} which provides a good fit to a broad
  variety of data, including the IR luminosity functions determined
  observationally by \citet{gruppioni13} at several redshifts up to
  $z\sim 4$ \citep[see also Figure~1 of][]{bonato14}.  The dot-dashed
  green histogram and the dot-dashed orange histogram show the
  contributions of strongly lensed (magnification, $\mu \ge 2$) and
  unlensed galaxies, respectively, while the black histogram shows the
  total.  The distribution of lensed galaxies was computed using the
  SISSA model \citep{lapi12}.  Although strongly lensed galaxies are a
  minor fraction of all galaxies with $S_{500}> 30\,$mJy, they become
  common at $z>4$ due to the combined effect of the increase with
  redshift of the optical depth to lensing and the magnification
  bias.  This will be addressed in a forthcoming paper, in which we
  present high-resolution ALMA imaging \citep{oteo16hires}.}
\label{fig:lensing}
\end{figure}

Comparison of the observed photometric redshift distribution for
our ultrared DSFGs with that expected by the \citet{bethermin15}
model (for sources selected with our flux density limits and
color criteria) reveals a significant mismatch, with the model
histogram skewed by $\delta z\approx 1$ bluewards of the observed
distribution.  This suggests that our current understanding of
galaxy evolution is incomplete, at least with regard to the most
distant, dust-enshrouded starbursts, plausibly because of the
influence of gravitational lensing, although the
\citeauthor{bethermin15} model does include a simple treatment of
this effect.  This issue will be addressed in a forthcoming paper
in which we present high-resolution ALMA imaging \citep[][see
also Fig.~\ref{fig:lensing}]{oteo16hires}.

The corresponding 8--1000-$\mu$m luminosities for our sample of
ultrared DSFGs, in the absence of gravitational lensing, range from
$5.0\times 10^{12}$ to $5.8\times 10^{13}$\,L$_\odot$, a median of
$1.3\times 10^{13}$\,L$_\odot$ and an interquartile range of
$9.7\times 10^{12}$ to $2.0\times 10^{13}$\,L$_\odot$.

Fig.~\ref{fig:lensing} demonstrates that the influence of
gravitational lensing cannot be wholly ignored.  Although
strongly lensed galaxies are a minor fraction of all galaxies
with $S_{500}> 30\,$mJy, they become more common at $z>4$ due to
the combined effect of the increase with redshift of the optical
depth to lensing and the magnification bias.  This will be
addressed in a forthcoming paper, in which we present
high-resolution ALMA imaging \citep{oteo16hires}.

\begin{figure}
\centerline{\psfig{file=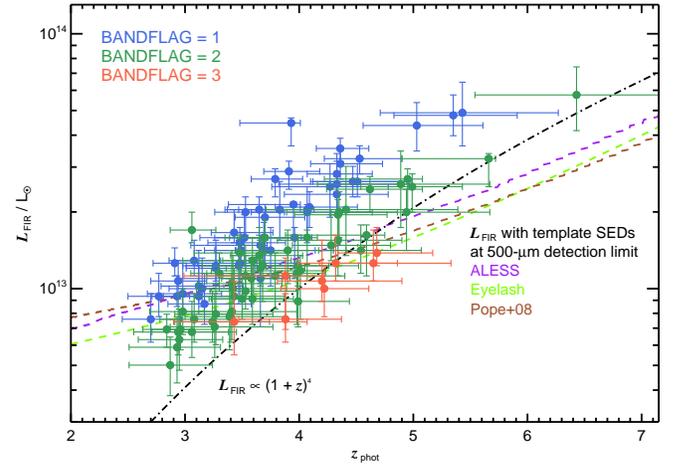,width=3.4in,angle=0}}
\caption{\lir\ as a function of $z_{\rm phot}$ for our sample,
  color-coded by {\sc bandflag}, with the $S_{500}>30$-mJy detection
  limits shown for our three best SED templates, and luminosity
  evolution of the form $\propto (1+z)^4$ illustrated. We see that the
  {\sc bandflag} $=1$, 2 and 3 galaxies lie in distinct regions, as
  one might expect.  The least luminous galaxies at any redshift are
  those detected only in the 500-$\mu$m filter, since in the SPIRE
  maps with the lowest spatial resolution they suffer considerably
  more flux boosting and blending.  The growing gap between the
  galaxies and the expected detection limits at $z>5$ is potentially
  interesting.}
\label{fig:lirz}
\end{figure}

In Fig.~\ref{fig:lirz} we show how the 8--1000-$\mu$m luminosities of
our ultrared DSFGs behave as a function of redshift, to help explain
the shape of our redshift distribution, and any biases.  The
$S_{500}>30$-mJy detection limit for our three best SED templates are
shown, as well as luminosity evolution of the form $(1+z)^4$, scaled
arbitrarily. The different {\sc bandflag} categories separate from one
another, as one might expect, where the least luminous galaxies at any
redshift are those detected only in the 500-$\mu$m filter, having
suffered considerably more flux boosting\footnote{{\sc bandflag} $=1$
  and 2 sources are extremely unlikely to coincide with positive noise
  peaks in two or three independent images simultaneously.} and
blending in the SPIRE maps with the lowest spatial resolution.  The
growing gap between the ultrared DSFGs and the expected detection
limits at $z>5$ is potentially interesting, possibly reflecting the
relatively low number of {\sc bandflag} $=3$ sources in our sample and
the growing influence of multi-band detections at the highest
redshifts.

\subsection{The space density of distant DSFGs}
\label{spacedensity}

With photometric redshift estimates for each of the sources in our
sample we can now set a lower limit on the space density, $\rho$, of
$S_{500}>30$-mJy ultrared DSFGs that lie at $z>4$. As summarised in
S\ref{nzsummary}, we find that $33\pm 6$\% of the sources in our sample lie
in the range $4<z<6$ and the space density of these DSFGs
is

\begin{equation}
   \label{eq:spacedensity}
\rho = \frac{N_z}{\mathcal{C} \, V_{\rm obs}} \times \frac{t_{\rm
    obs}}{t_{\rm burst}} \,\,\,\,\,\,\, {\rm Mpc}^{-3},
\end{equation}

\noindent
where $N_z$ represents the number of sources within $4<z<6$, $V_{\rm
  obs}$ is the comoving volume contained within the redshift range
considered, $t_{\rm obs}/t_{\rm burst}$ is a duty-cycle correction,
since the ongoing, obscured starburst in DSFGs has a finite duration,
where $t_{\rm burst}\approx 100$\,Myr is in agreement with their
expected gas depletion times \citep{ivison11, bothwell13} but is
uncertain at the $\approx 2\times$ level.  $\mathcal{C}$ is the
completeness correction required for our sample, as discussed at
length in \S\ref{parentsample}--\ref{summarycompleteness}.  $V_{\rm
  obs}$ is the comoving volume contained within $4<z<6$, given by

\begin{equation}
   \label{eq:surveyvolume}
   V_{\rm obs} = \frac{4\pi}{3} \int_{z=4}^{z=6} \frac{c/H_{\rm
       0}}{\sqrt{\Omega_{\rm M} (1+z')^3 + \Omega_{\rm V}}} dz' \,\,\,\,\,\,\,
   {\rm Mpc}^3
\end{equation}
\citep{hogg99}, which we scale by the fractional area of sky that was
surveyed, $\approx 600$\,deg$^2$, or $\approx 1.5$\%.

Applying these corrections we estimate that ultrared, DSFGs at $z>4$
have a space density of $\approx 6\times 10^{-7}$\,Mpc$^{-3}$.  Our
work represents the first direct measurement of the space density of
$z>4$ DSFGs at such faint flux-density limits and as such it is not
possible to make a direct comparison with previous studies in the
literature.  For example, \citet{asboth16} recently presented the
number counts of ultrared, 500--$\mu$m--selected DSFGs, identified in
the 274-deg$^2$ HerMES Large Mode Survey (HELMS).  However, the
\citeauthor{asboth16} galaxies are considerably brighter than ours,
meaning a significant fraction will be gravitationally lensed, and
they lack redshift estimates, so it is impossible to judge
meaningfully whether their source density is consistent with the
results presented here.

\subsection{Relationship of DSFGs with other galaxy populations}

It has been suggested by a number of authors \citep[e.g.][]{simpson14,
  toft14,ikarashi15} that high-redshift DSFGs may be the progenitors
of the population of massive, quiescent galaxies that have been
uncovered in near-IR surveys \citep[e.g.][]{vandokkum08, newman12,
  krogager14, straatman14}. These galaxies are generally found to be
extremely compact which, when taken in conjunction with their high
stellar masses, $\approx 10^{11}$\,M$_{\odot}$, and high redshifts,
$z\gs 2$, motivates the idea that the stellar component was formed
largely during an intense starburst phase, enshrouded in dust.

Is the comoving space density of ultrared, high-redshift DSFGs
consistent with that of massive, high-redshift, quiescent galaxies? As
discussed earlier, the $4<z<6$ DSFGs presented in this work have a
comoving space density of $\approx 6 \times 10^{-7}$\,Mpc$^{-3}$. As a
comparison, we use the galaxies in the sample presented by
\citet{straatman14}, which were classified as quiescent via $UVJ$
selection \citep[e.g.][]{labbe05} and are drawn from a mass-limited
sample ($>4\times 10^{10}$\,M$_{\odot}$). These galaxies were selected
to lie in the redshift range $3.4<z<4.2$ and were estimated to have a
median stellar age of $\approx 0.8$\,Gyr, indicating a typical
formation epoch of $z\approx 5$, making them an ideal match to our
sample of $4<z<6$ DSFGs.

The quiescent sources presented by \citet{straatman14} have a comoving
space density of $\approx 2\times 10^{-5}$\,Mpc$^{-3}$, $\approx
30\times$ more numerous than the sample of DSFGs presented here.  Even
at $M_{\rm stars}\gs 10^{11}$\,M$_\odot$, \citeauthor{straatman14}
estimate a space density of $\approx 4\times 10^{-6}$\,Mpc$^{-3}$ for
their quiescent near-IR galaxies, still almost an order of magnitude
higher than our $z>4$ DSFGs.  This indicates clearly that $z>4$ DSFGs
cannot account for the formation of massive, quiscent galaxies at
$z\sim 3$--4 when selected at the flux-density levels we have been
able to probe with {\it Herschel}.  Even an infeasibly short duration
of $\ls 10$\,Myr for the starburst phase of DSFGs is insufficient to
bring the comoving space densities of the two populations into
agreement, except at the very highest masses.  Instead, our
$S_{350}\approx S_{500}\approx 30$-mJy flux density limits are
selecting the rarest, most FIR-bright objects on the sky ---
hyperluminous galaxies \citep[e.g.][]{fu13,ivison13} --- which can
form a galaxy with $\gs10^{11}$\,M$_{\odot}$ of stars in $\ls
100$\,Myr, and/or less massive galaxies caught during a tremendously
violent, short-lived phase, or gravitationally magnified by a chance
alignment, populations that --- even collectively --- are considerably
rarer than massive, high-redshift, quiescent galaxies.

The ALMACAL program of \citet{oteo16almacal} has shown that that
$S_{870}\gs 1$-mJy DSFGs with SFRs of $\approx
50$--100\,M$_\odot$\,yr$^{-1}$ are three orders of magnitude more
common than our $z>4$ {\it Herschel}-selected DSFGs, such that
$\approx 1$--2\% of them lying at $z>4$ may account for the massive,
quiescent near-IR-selected galaxies. Given the limited mapping speed
of ALMA, even this fainter, more numerous DSFG population will be best
accessed via a facility designed to obtain deep, wide-field imaging in
passbands spanning 350\,$\mu$m through 2\,mm, either a large dish or a
compact array equipped with focal-plane arrays.

We must therefore admit that although the progenitors of the most
massive ($\gs10^{11}$-M$_\odot$) quiescent galaxies are perhaps just
within our grasp, if we can push this color-selection technique
further, the progenitors of the more general near-IR-selected
quiescent galaxy population lie below the flux-density regime probed
directly by {\it Herschel}.  The progenitors of $z>6$ quasars,
discussed in \S\ref{intro}, remain similarly elusive: our ultrared
DSFG space density is well matched, but we have yet to unveil any of
the $z>6$ galaxies that may be hidden within our sample.

\section{Conclusions}
\label{conclusions}

We have presented follow up SCUBA-2 and LABOCA imaging of a sample of
109 ultrared DSFGs with {\it Herschel} SPIRE colors of
$S_{500}/S_{250} \geq 1.5$ and $S_{500}/S_{350} \geq 0.85$, thereby
improving the accuracy of FIR-/submm-based photometric redshifts.
After selecting the three SED templates most suitable for determining
photometric redshifts, from a parent sample of seven, we performed two
further sanity checks, looking for significant systematics and finding
none, suggesting a high degree of accuracy.  We then determine a
median redshift, $\hat{z}_{\rm phot}=3.66$, and an interquartile range
of $z_{\rm phot}=3.30$--4.27, with a median rest-frame 8--1000-$\mu$m
luminosity, $\hat{L}_{\rm IR}=1.3\times 10^{13}$\,L$_\odot$.  We
determine that $32\pm 5$\% lie at $z_{\rm phot}>4$, and that the space
density of such galaxies is $\approx 6 \times 10^{-7}$\,Mpc$^{-3}$.

Comparison of the observed photometric redshift distribution for our
ultrared DSFGs with that expected by a phenomenological model of
galaxy evolution reveals a significant mismatch, with the model skewed
by $\delta z\approx 1$ bluewards of the observed redshift
distribution.

Although the progenitors of the most massive ($\gs10^{11}$-M$_\odot$)
near-IR-selected quiescent galaxies are perhaps just within our grasp,
if we push this color-selection technique further, the progenitors of
the more general near-IR-selected quiescent galaxy population lie
below the flux-density regime probed directly by {\it Herschel}.  Our
ultrared DSFG space density is relatively well matched to that of
$z>6$ quasars, but their progenitors remain elusive since we have yet
to unveil any $z>6$ galaxies in our sample.

With this unique sample, we have substantially increased the number of
$z>4$ dusty galaxies, partially fulfilling the promise of early
predictions for the negative $K$ correction in the submm band
\citep{blain93}.  However, although we can claim considerable success
in significantly enlarging the known sample of ultrared DSFGs at
$z>4$, we must acknowledge that over half of our sources lie at
$z<4$. Because of this, and the uncertain fraction of spurious sources
in our parent ultrared DSFG catalog, we regard further refinement of
the ultrared selection technique as both possible and necessary.

Finally, we draw attention to an interesting source,
HATLAS\,J004223.5$-$334340 (SGP-354388), which we have dubbed the
`Great Red Hope' (or GRH).  This system is resolved in our LABOCA and
SCUBA-2 imaging, with a total 850-$\mu$m [870-$\mu$m] flux density of
$58\pm7$ [$64\pm11$]\,mJy. In a 3-mm continuum map from ALMA covering
$\approx 1$\,arcmin$^2$ \citep{fudamoto16}, we see a number of
discrete DSFGs \citep{oteo16grh}, most of which display a single
emission line\footnote{Redshifts of 3.7, 4.9, 6.0 or 7.2 are
  plausible, if this line is due to CO $J=4$--3, $J=5$--4, $J=6$--5 or
  $J=7$--6.} at 98.4\,GHz, an overdensity of galaxies that continues
on larger scales, as probed by wide-field LABOCA imaging
\citep{lewis16}.  Photometric redshifts are challenging under these
circumstances, given the confusion in the {\it Herschel} bands.  Using
the \citeauthor{swinbank14} ALESS SED template with point-source flux
densities suggests the lowest plausible redshift for these galaxies is
$\approx 4.0$, while the method used throughout this work to measure
flux densities gives $z_{\rm phot}\sim 5.4$. At anything like this
distance, this is a remarkable cluster of ultrared DSFGs.

\section*{Acknowledgements}

RJI, AJRL, VA, LD, SM and IO acknowledge support from the European
Research Council (ERC) in the form of Advanced Grant, 321302, {\sc
  cosmicism}.  HD acknowledges financial support from the Spanish
Ministry of Economy and Competitiveness (MINECO) under the 2014
Ram\'{o}n y Cajal program, MINECO RYC-2014-15686.  IRS acknowledges
support from the Science and Technology Facilities Council (STFC,
ST/L00075X/1), the ERC Advanced Grant, DUSTYGAL 321334, and a Royal
Society/Wolfson Merit Award. DR acknowledges support from the National
Science Foundation under grant number AST-1614213 to Cornell
University. MN acknowledges financial support from the Horizon 2020
research and innovation programme under Marie Sklodowska-Curie grant
agreement, 707601.  The {\it Herschel}-ATLAS is a project with {\it
  Herschel}, which is an ESA space observatory with science
instruments provided by European-led Principal Investigator consortia
and with important participation from NASA. The {\it H}-ATLAS website
is {\tt www.h-atlas.org}. US participants in H-ATLAS acknowledge
support from NASA through a contract from JPL. The JCMT is operated by
the East Asian Observatory on behalf of The National Astronomical
Observatory of Japan, Academia Sinica Institute of Astronomy and
Astrophysics, the Korea Astronomy and Space Science Institute, the
National Astronomical Observatories of China and the Chinese Academy
of Sciences (Grant No.~XDB09000000), with additional funding support
from STFC and participating universities in the UK and Canada; Program
IDs: M12AU24, M12BU23, M13BU03, M12AN11, M13AN02. Based on
observations made with APEX under Program IDs: 191A-0748,
M.090.F-0025-2012, M.091.F-0021-2013, M-092.F-0015-2013,
M-093.F-0011-2014.\smallskip

{\it Facilities:} \facility{JCMT, APEX, {\it Herschel}.}

\bibliographystyle{apj}
\bibliography{rji}

\appendix

In this Appendix we present the {\it Herschel} SPIRE, JCMT/SCUBA-2 and
APEX/LABOCA imaging of our red galaxy sample in the GAMA 9-hr, 12-hr
and 15-hr fields, as well as the NGP and SGP fields. In each column,
from left to right, we show 250-, 350-, 500- and 850-$\mu$m
[870-$\mu$m for LABOCA] cut-out images, each $3^\prime\times 3^\prime$
and centered on the (labelled) galaxy. The 250- and 850-$\mu$m
[870-$\mu$m] cut-out images have been convolved with
7$^{\prime\prime}$ and 13$^{\prime\prime}$ [19$^{\prime\prime}$]
Gaussians, respectively. The 45$^{\prime\prime}$ aperture used to
measure $S_{\rm tot}$ is shown. A 60$^{\prime\prime}$ aperture was
also used but is not shown, to aid clarity. The annulus used to
measure the background level is shown in the uppermost case (this is
correspondingly larger for the 60$^{\prime\prime}$ aperture -- see
\S\ref{scuba2fluxes}). SPIRE images are displayed from $-6$ to
+60\,mJy\,beam$^{-1}$; SCUBA and LABOCA images are displayed from $-3$
to +30\,mJy\,beam$^{-1}$; both scales are relative to the local
median. North is up and East is left.

\setcounter{figure}{0}
\begin{figure*}
\centerline{\psfig{file=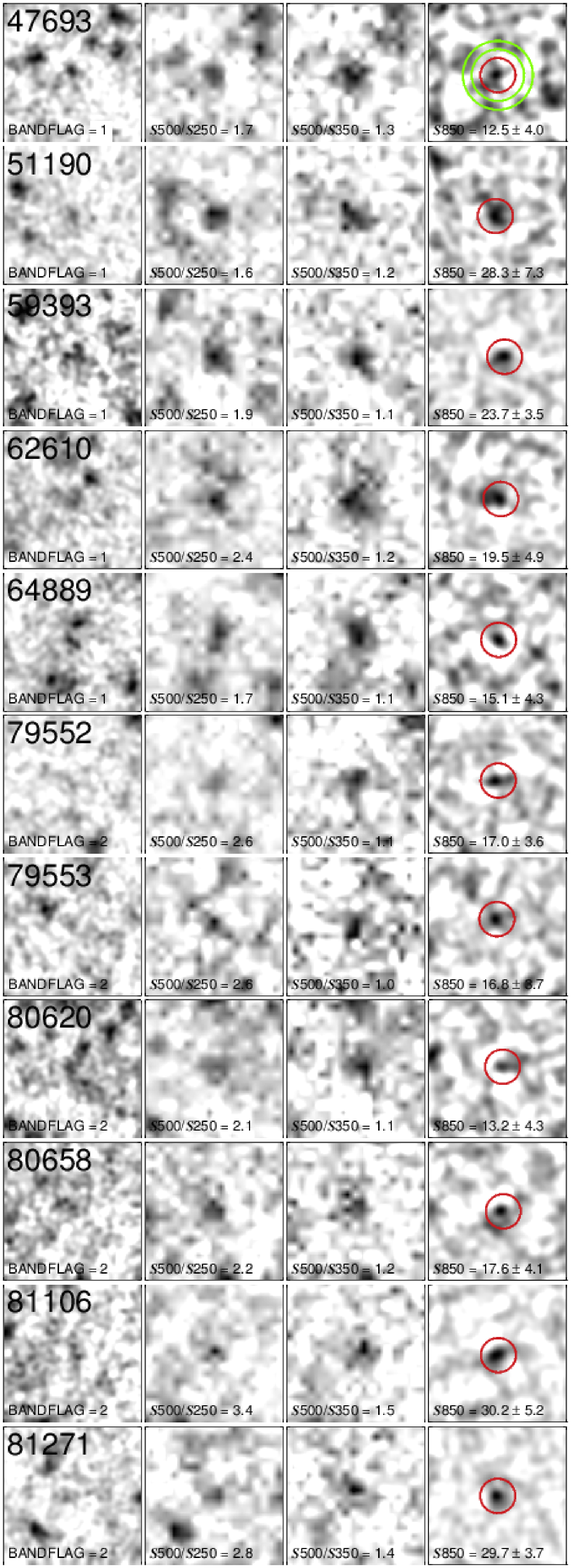,width=3.5in,angle=0}
\hspace{0.1cm}
\psfig{file=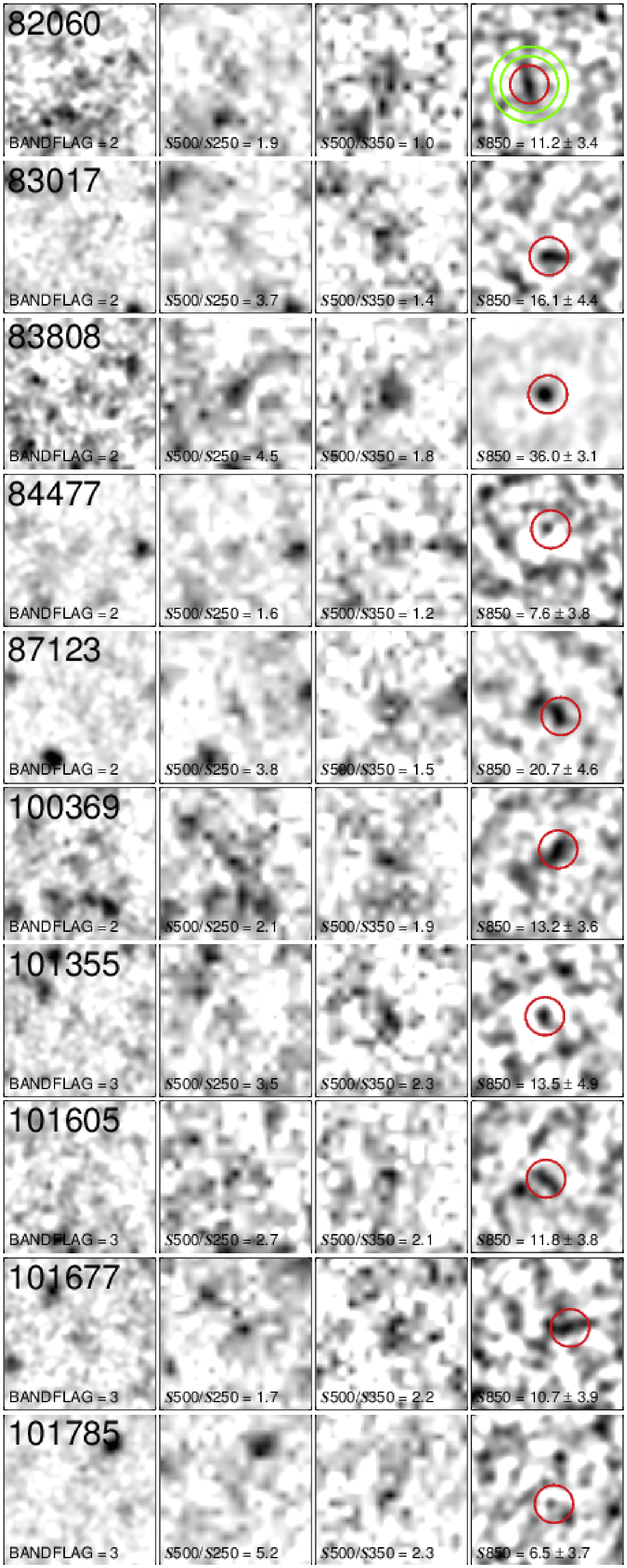,width=3.5in,angle=0}}
\caption{Targets in the GAMA 9-hr field, observed by SCUBA-2.}
\label{fig:gama09}
\end{figure*}

\begin{figure*}
\centerline{\psfig{file=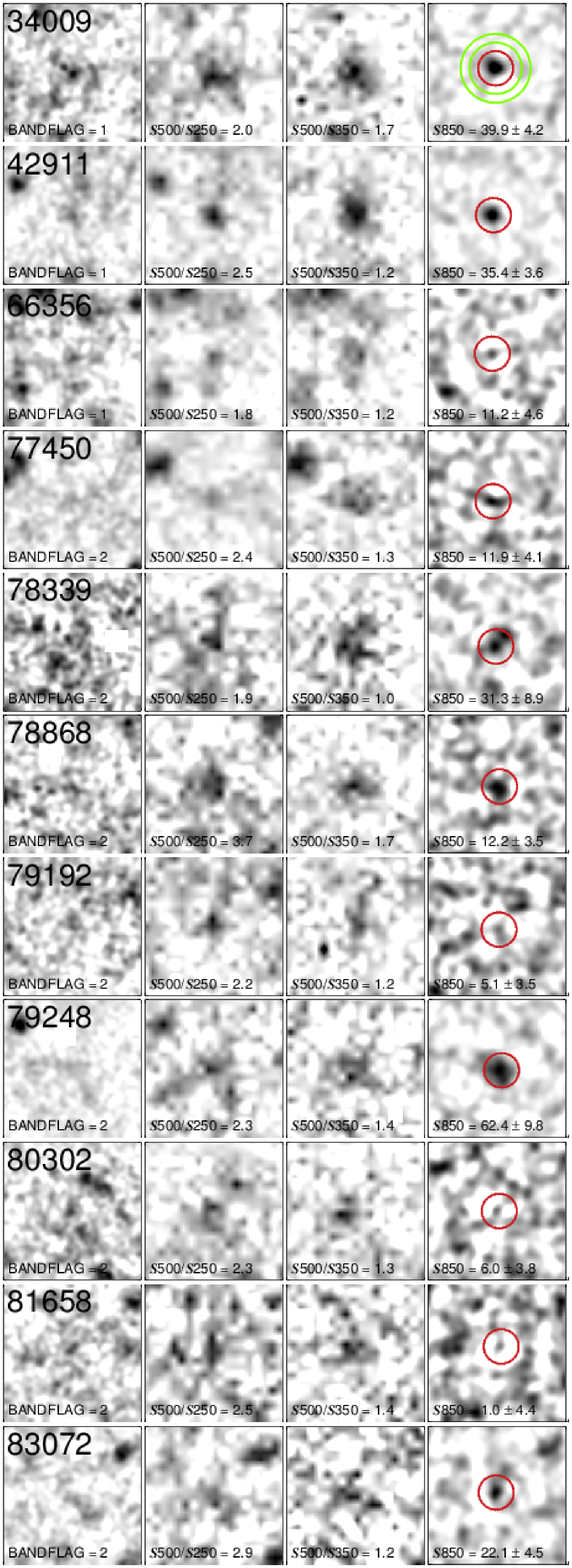,width=3.5in,angle=0}
\hspace{0.1cm}
\psfig{file=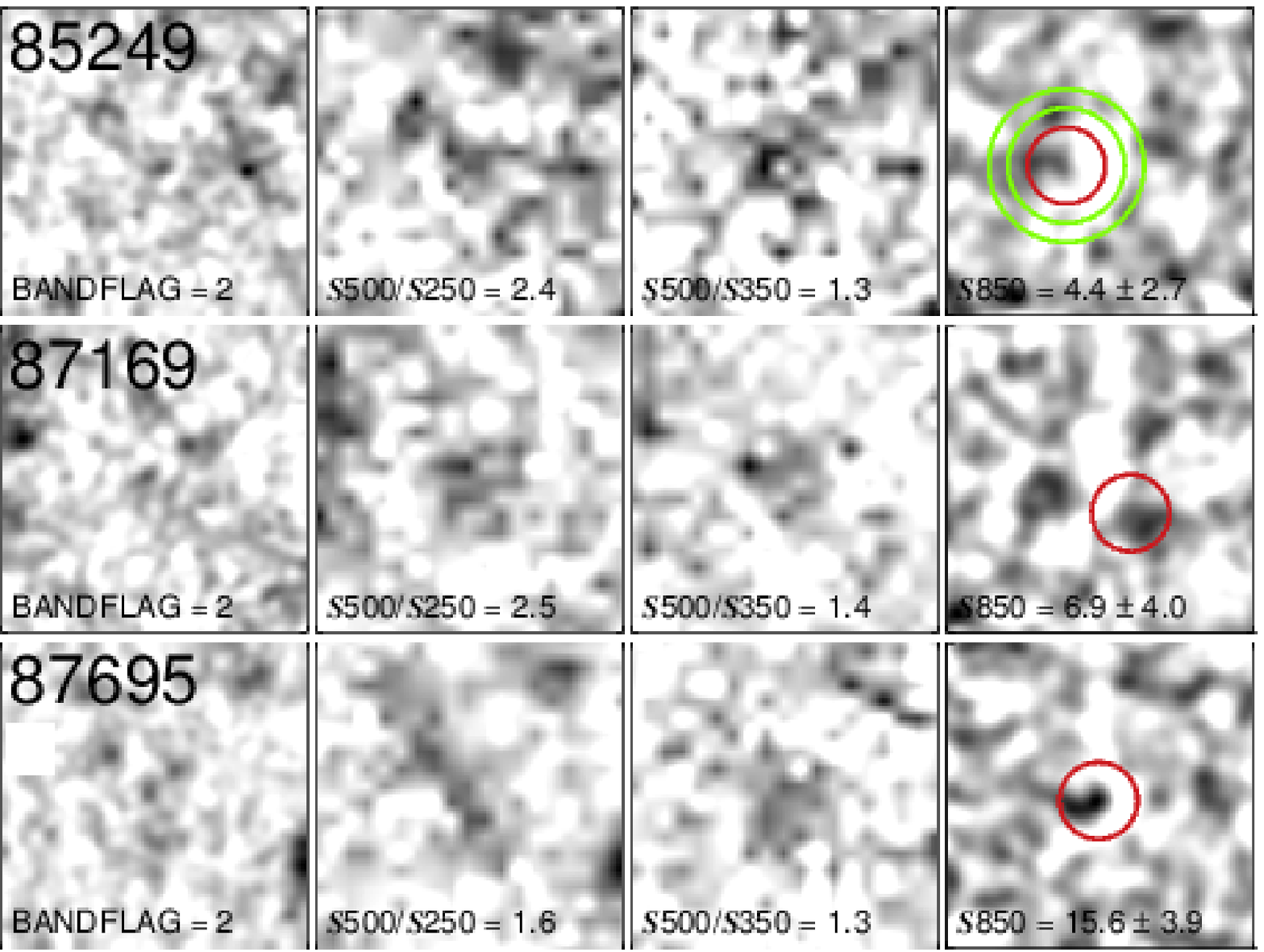,width=3.5in,angle=0}}
\caption{Targets in the GAMA 12-hr field, observed by SCUBA-2.}
\label{fig:gama12}
\end{figure*}

\begin{figure*}
\centerline{\psfig{file=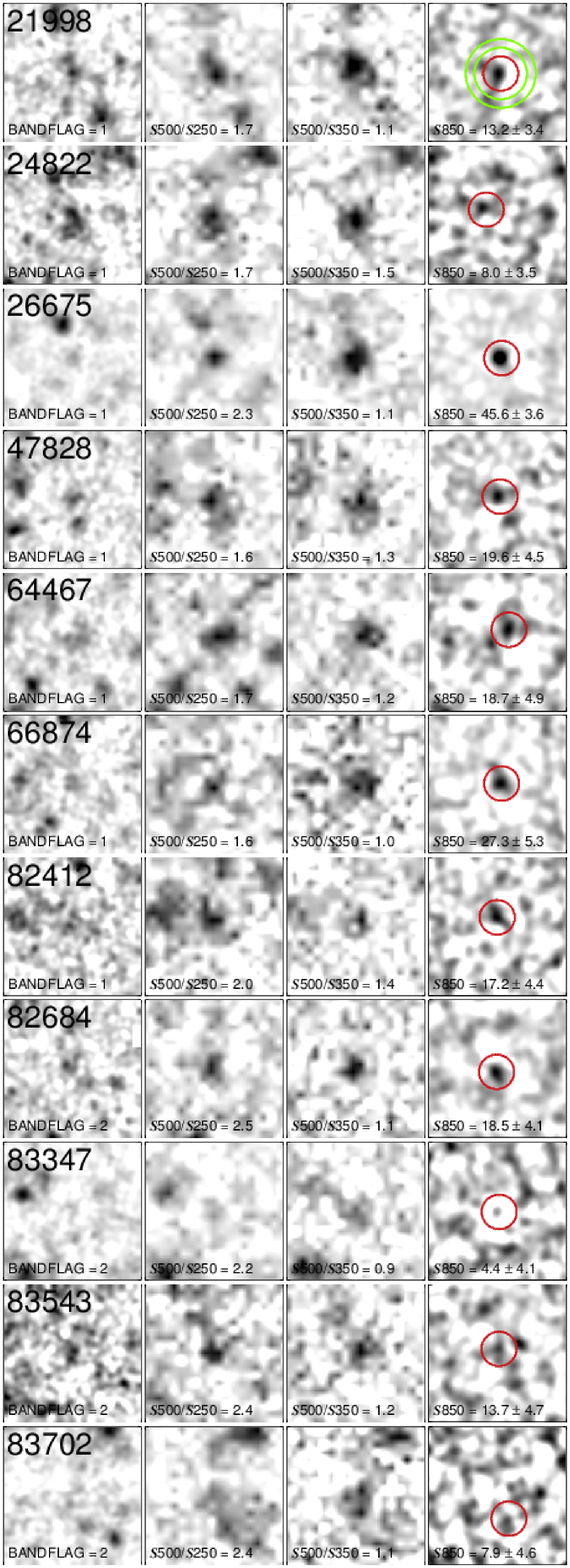,width=3.5in,angle=0}
\hspace{0.1cm}
\psfig{file=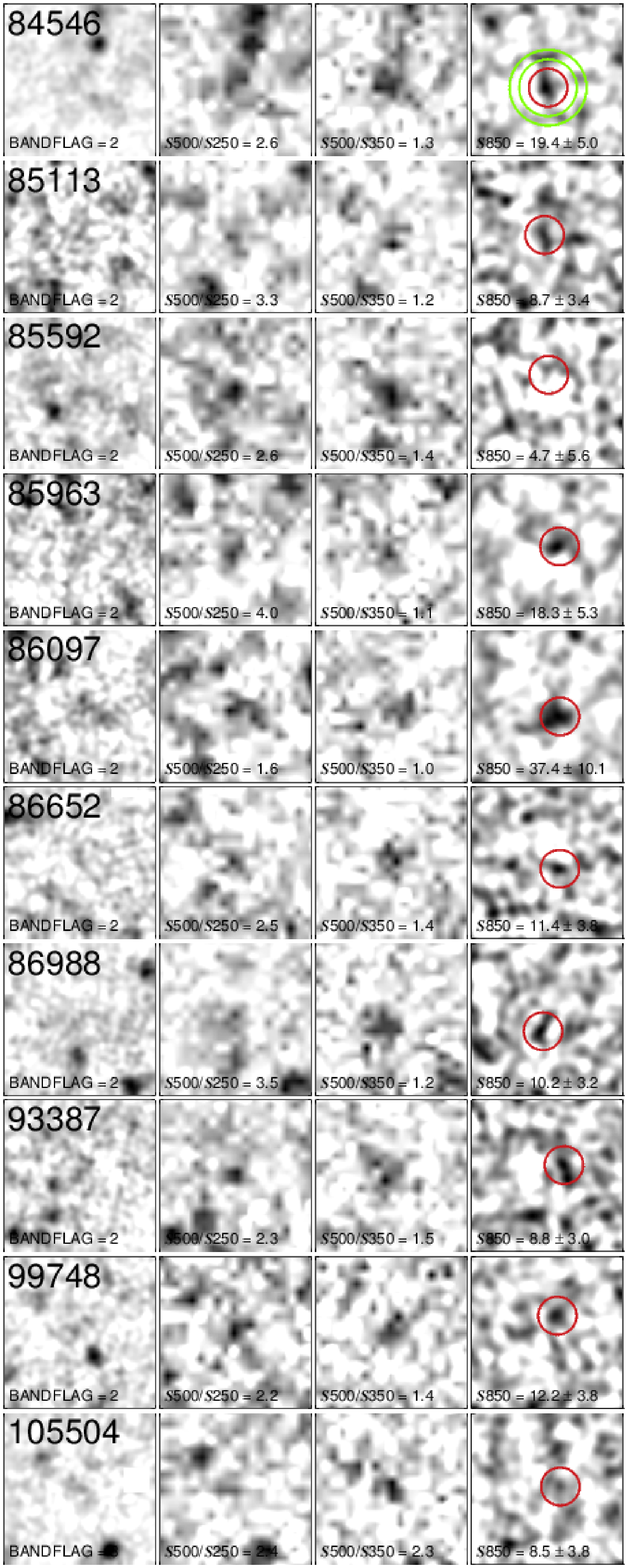,width=3.5in,angle=0}}
\caption{Targets in the GAMA 15-hr field, observed by SCUBA-2.}
\label{fig:gama15}
\end{figure*}

\begin{figure*}
\centerline{\psfig{file=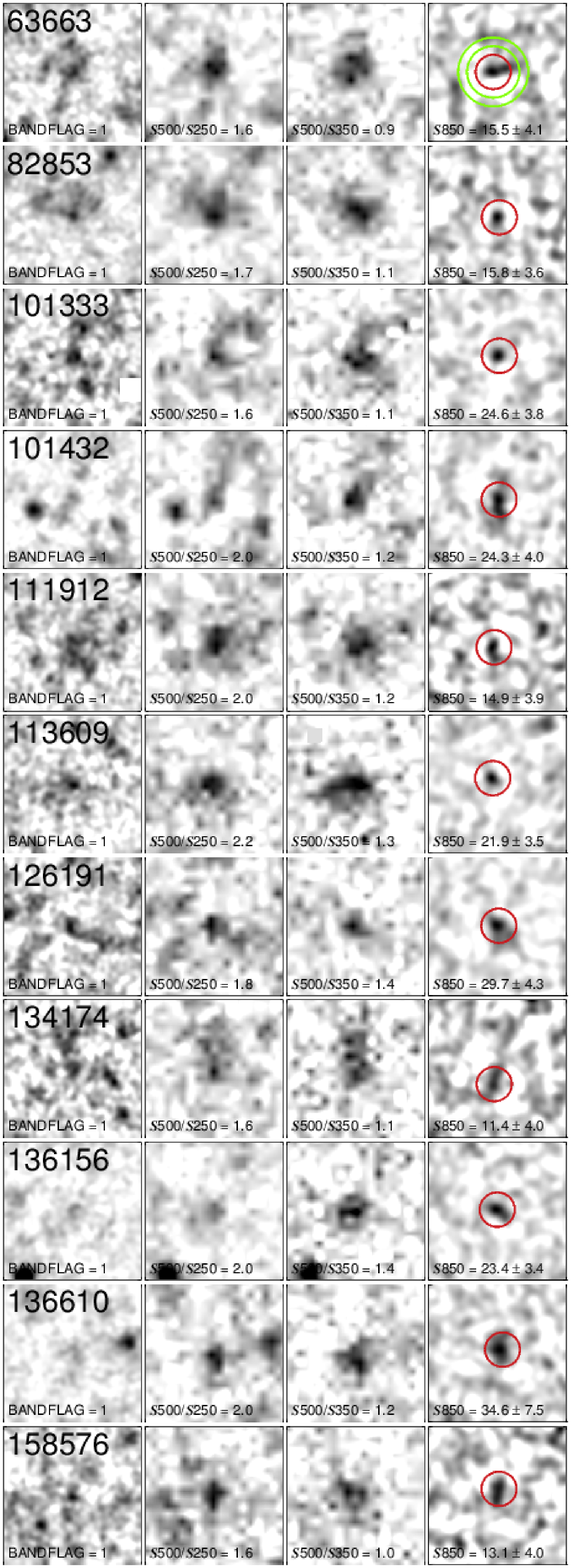,width=3.5in,angle=0}
\hspace{0.1cm}
\psfig{file=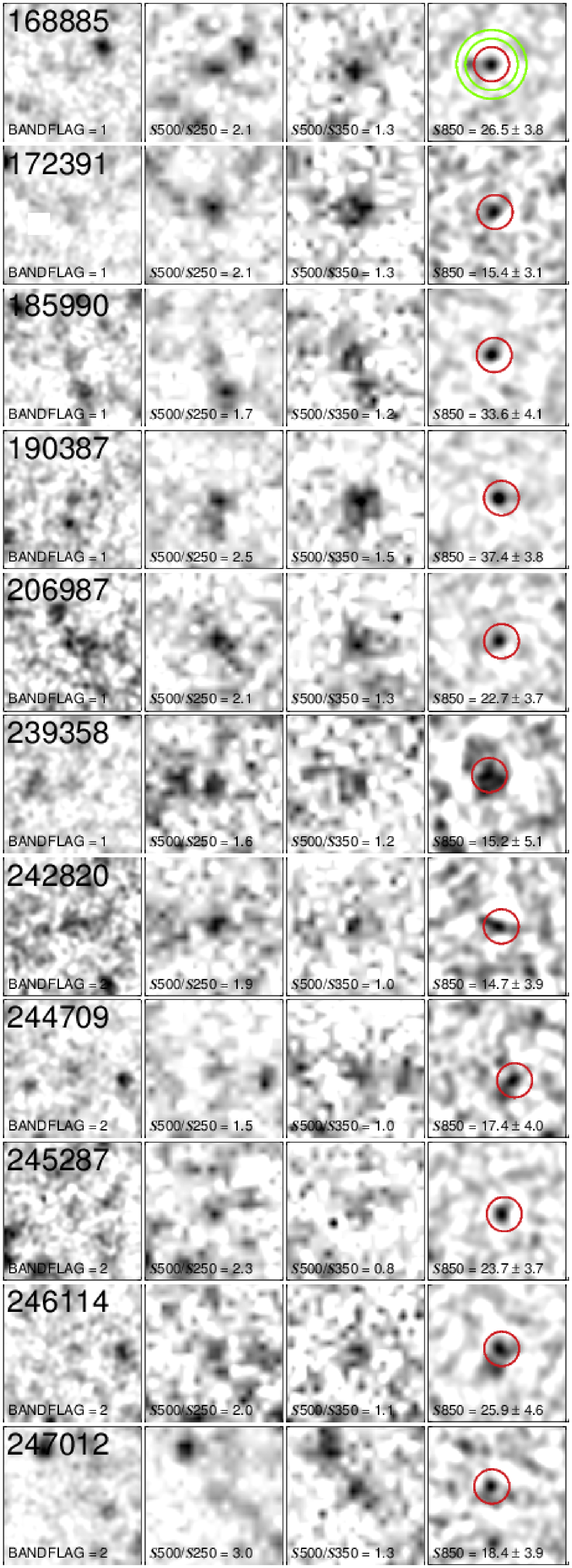,width=3.5in,angle=0}}
\caption{Targets in the NGP field, observed by SCUBA-2.}
\label{fig:ngp}
\end{figure*}

\setcounter{figure}{3}
\begin{figure*}
\centerline{\psfig{file=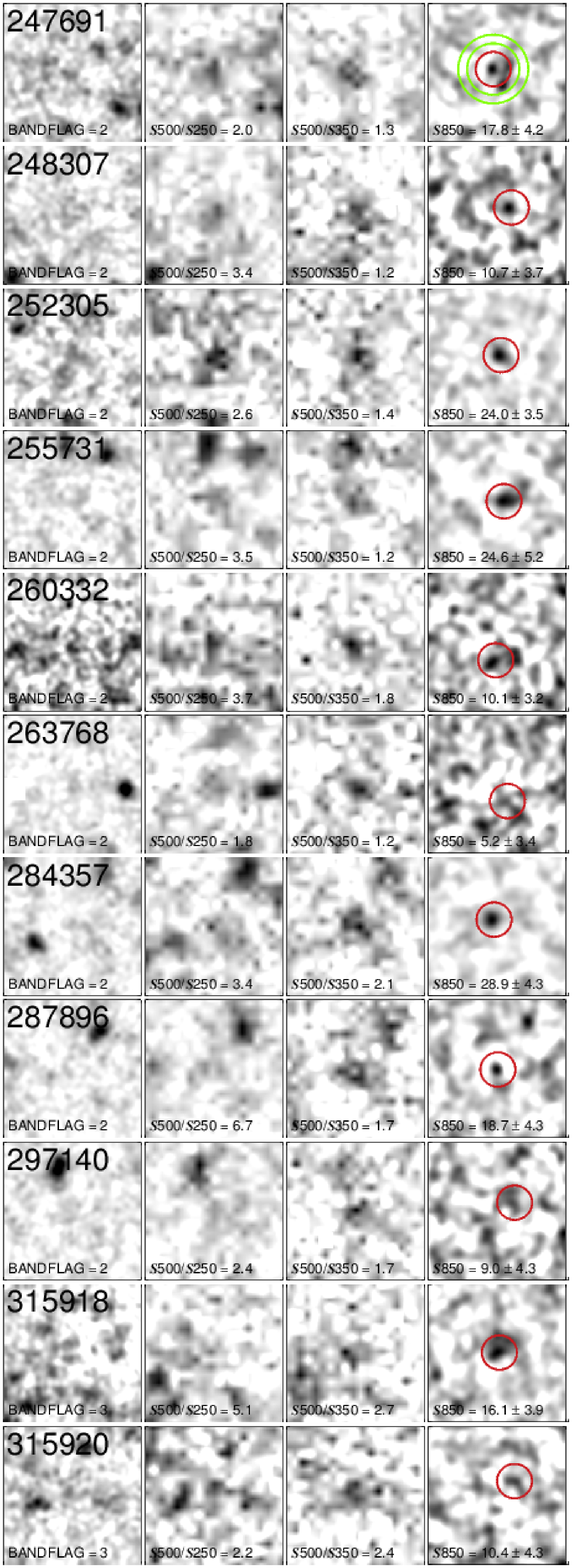,width=3.5in,angle=0}
\hspace{0.1cm}
\psfig{file=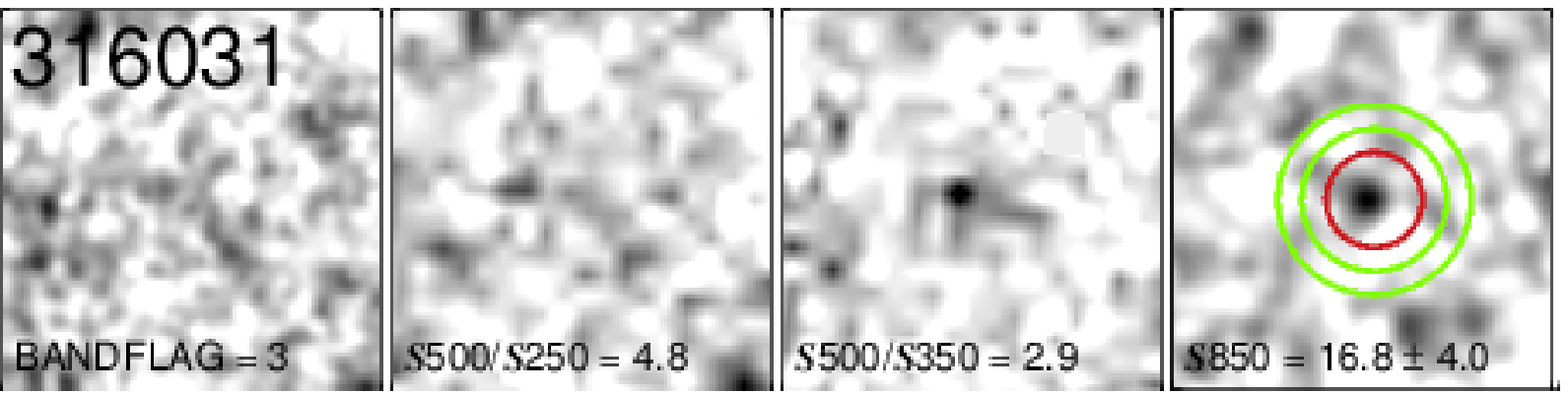,width=3.5in,angle=0}}
\caption{Cont...}
\end{figure*}

\begin{figure*}
\centerline{\psfig{file=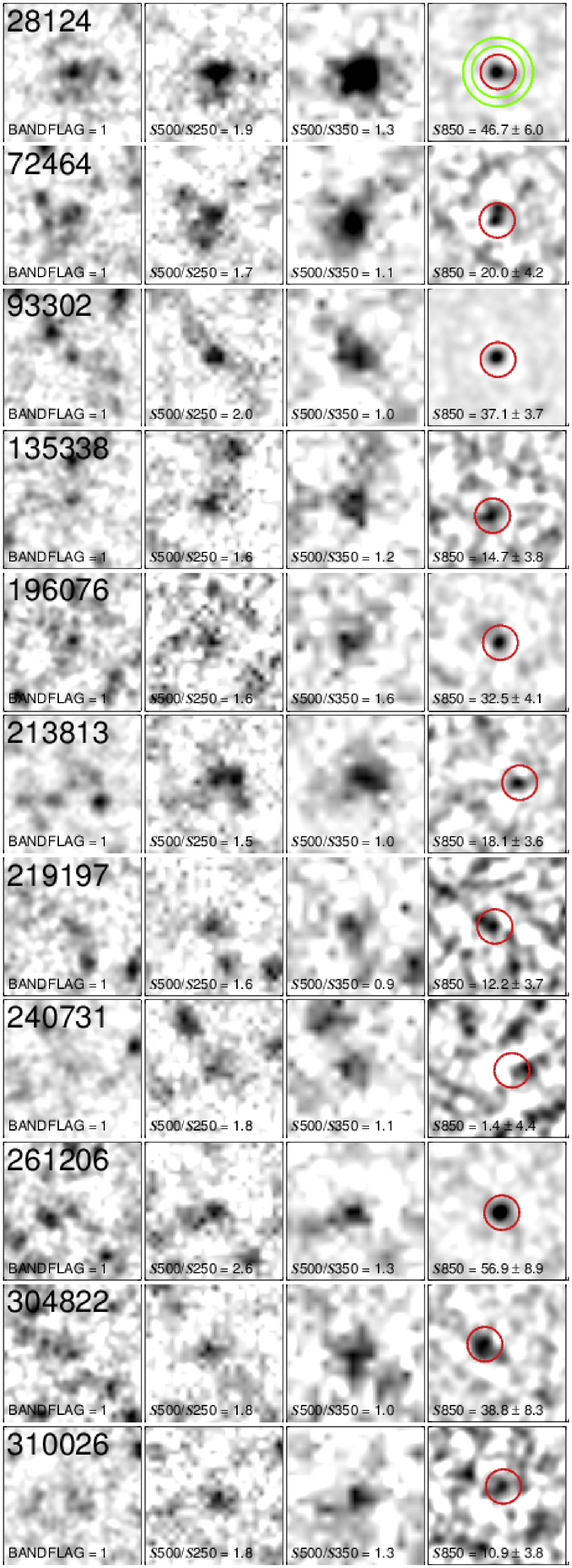,width=3.5in,angle=0}
\hspace{0.1cm}
\psfig{file=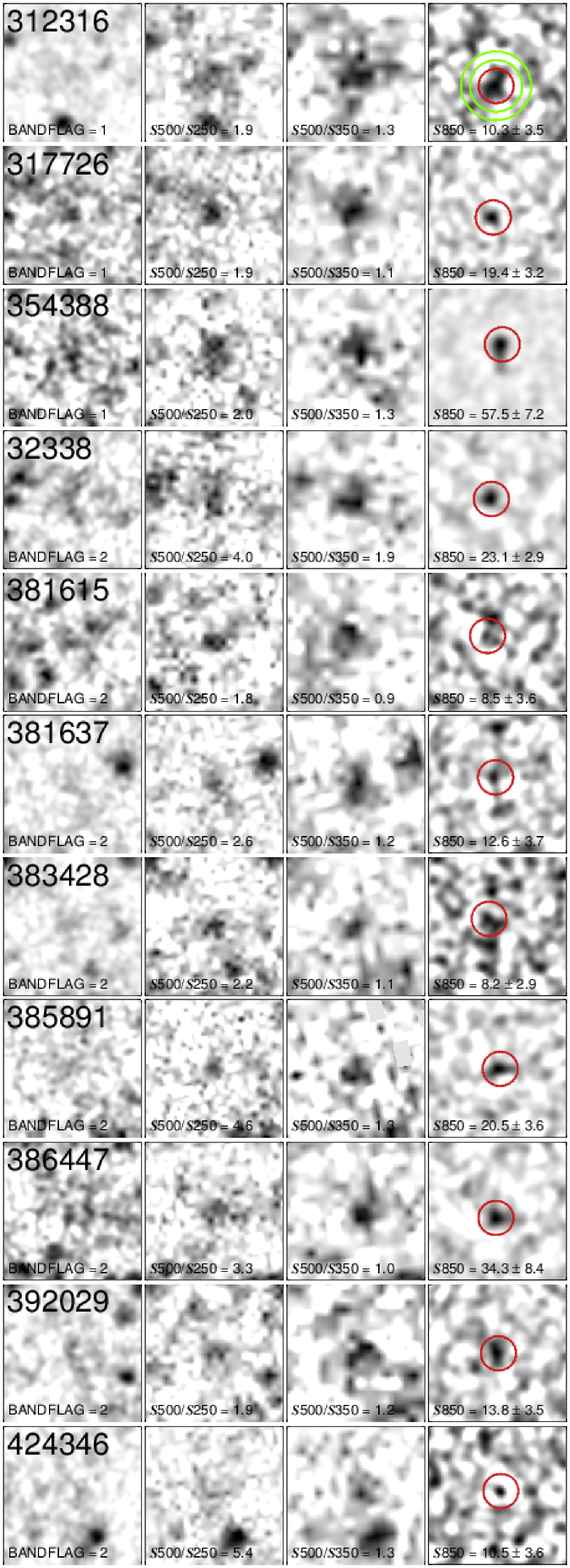,width=3.5in,angle=0}}
\caption{Targets in the SGP field, observed by SCUBA-2.}
\label{fig:sgp}
\end{figure*}

\setcounter{figure}{4}
\begin{figure*}
\centerline{\psfig{file=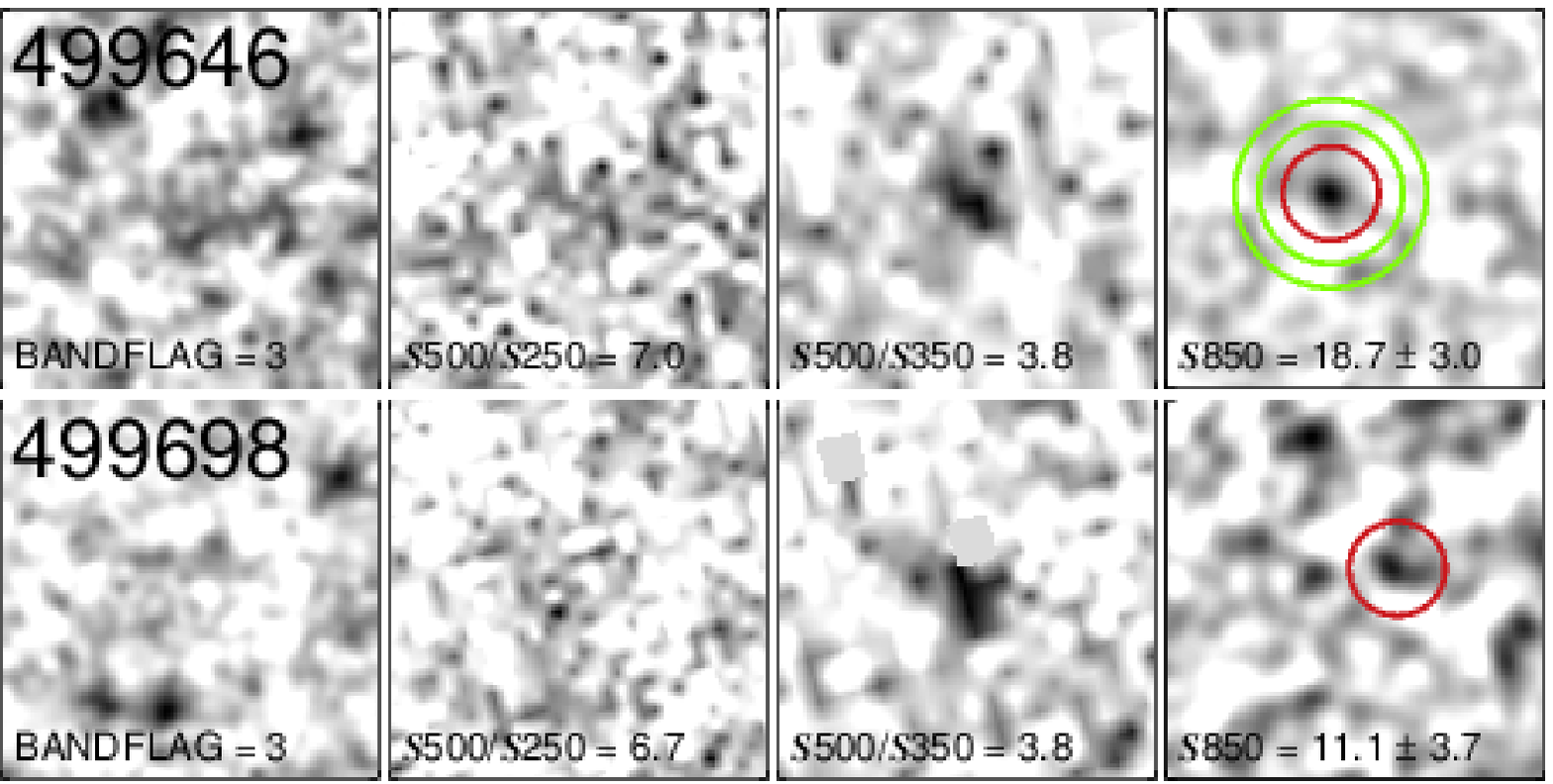,width=3.5in,angle=0}
\hspace{0.1cm}
\psfig{file=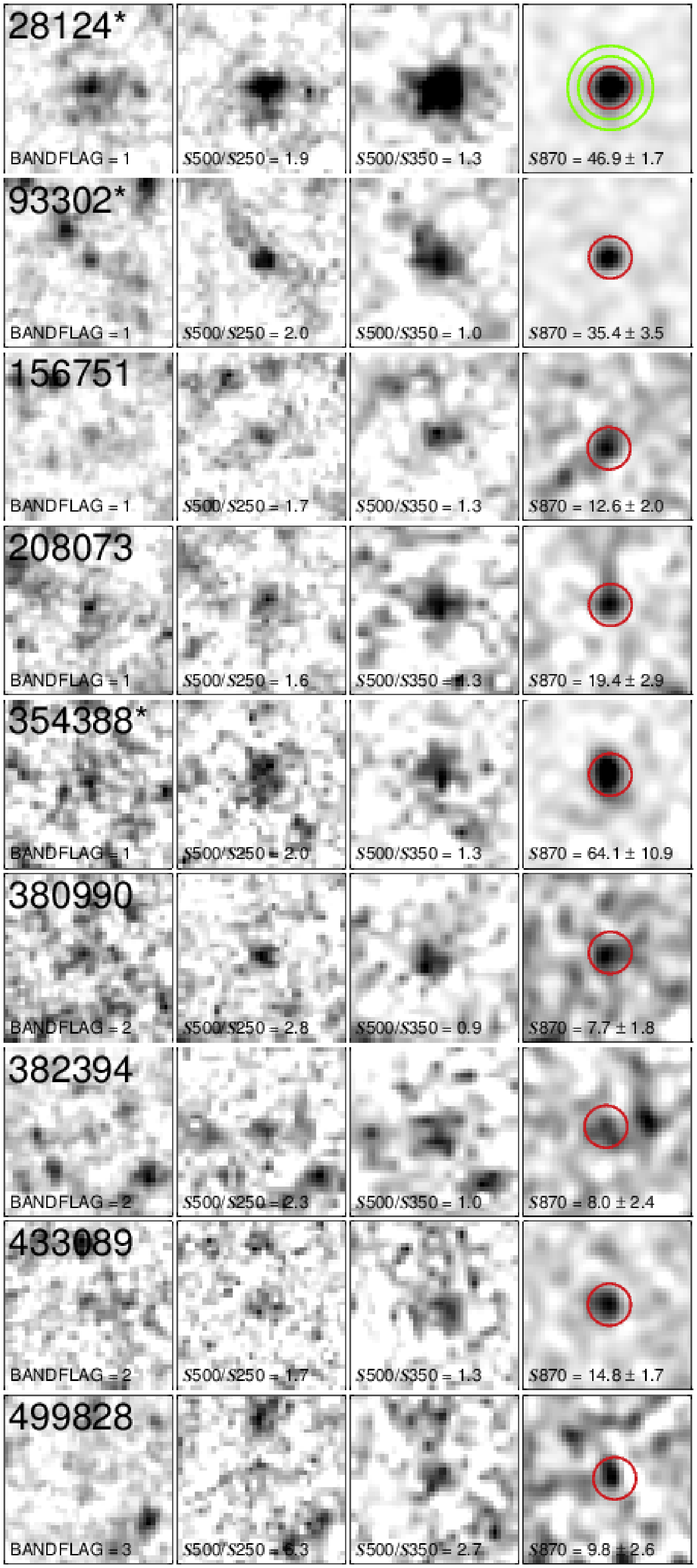,width=3.5in,angle=0}}
\caption{\textit{Left:} Cont... \textit{Right:} Targets in the SGP field, observed by LABOCA.}
\end{figure*}

\end{document}